\documentclass[12pt, prd, aps, superscriptaddress, notitlepage, preprint]{revtex4-2}

\usepackage{natbib}
\usepackage{mathrsfs}

\usepackage[utf8]{inputenc}
\usepackage{amsfonts,bm}
\usepackage{amsmath,amssymb,graphicx,textcomp,xcolor}
\usepackage{hyperref}
\usepackage[caption=false]{subfig}
\usepackage{slashed}
\usepackage{comment}
\usepackage{cleveref}

\begin{document}

\newcommand*{\MIT}{Center for Theoretical Physics -- A Leinweber Institute, Massachusetts Institute of Technology, Cambridge, MA 02139, USA}\affiliation{\MIT}
\newcommand*{\CU}{Physics Department, Columbia University, New York City, New York 10027, USA}\affiliation{\CU}

\newcommand{\msbar}{$\overline{\rm MS}$ }
\newcommand{\kl}{K_{\textrm{L}}}
\newcommand{\hw}{H_{\textrm{W}}}
\newcommand{\chw}{\mathcal{H}_{\textrm{W}}}
\newcommand{\lla}{\left\langle}
\newcommand{\rra}{\right\rangle}
\newcommand{\tr}{\mathrm{Tr}}
\newcommand{\sinc}{\sin_{\textrm{c}}}
\newcommand{\sumint}[1]{\,\,\,
\mathclap{\displaystyle\int}\mathclap{\textstyle\sum}\,\,\,{}_{#1}}
\newcommand{\gfive}{\Gamma^5}
\newcommand{\gL}{\Gamma^L}
\newcommand{\tsep}{t_{\textrm{sep}}}
\newcommand{\dmax}{\delta_{\rm{max}}}
\newcommand{\wh}{\widehat}
\newcommand{\wt}{\widetilde}
\newcommand{\Res}{\textrm{Res}}
\newcommand{\gf}{G_{\rm F}}
\newcommand{\nf}{N_{\rm f}}
\newcommand{\lat}{\textrm{lat}}

\title{Renormalizing a three-flavor lattice calculation of the two-photon contribution to $K_L\to\mu^+\mu^-$}
\author{En-Hung Chao}\affiliation{\MIT}
\author{Norman Christ}\affiliation{\CU}
\author{Ceran Hu}\affiliation{\CU}
\preprint{MIT-CTP/6068}
\date{July 12, 2026}

\begin{abstract}
The Standard Model prediction for the rare $K_L\to\mu^+\mu^-$ decay depends critically on the long-distance contribution coming from the exchange of two photons.  Such a contribution can be computed using lattice QCD and an effective three-flavor theory including only the $u$, $d$ and $s$ quarks, provided terms falling as the inverse square of the omitted charm quark mass, $1/m_c^2$ are neglected.  Because of the missing Glashow-Iliopoulos-Maiani cancelation, this three-flavor theory contains additional low-energy constants that depend on $m_c$.  Here we show how these constants can be determined from a practical four-flavor lattice QCD calculation performed on a small volume with $u$ and $d$ quark masses that are heavier than physical.
\end{abstract}

\maketitle

\tableofcontents

\newpage

\section{Introduction}

The rare $K_L\to\mu^+\mu^-$ decay is a strangeness-changing neutral-current process which cannot occur at tree level in Standard Model.  This process depends on short-distance dominated, one-loop effects involving multiple $W$ and/or $Z$ exchanges and is therefore expected to show increased sensitively to new phenomena that are not described by the Standard Model. It is of special interest because the decay rate has been measured to 1.6\% accuracy~\cite{E871:2000wvm} in the E871 Brookhaven experiment and the short-distance, one-loop Standard Model prediction has also been computed to NLO in QCD perturbation theory~\cite{Buchalla:1995vs, Gorbahn:2006bm}.  

A direct comparison between these two results is not yet possible because of a two-loop electroweak process in which a two-photon intermediate state joins the weakly decaying kaon and the two-muon final state.  The imaginary part of this two-loop amplitude is known from the $K_L\to\gamma\gamma$ decay rate and the optical theorem.  However, the real part of this amplitude is expected to be of the same size as the one-loop Standard Model prediction and must be determined before this decay can be used for an important test of the short-distance, one-loop predictions of the Standard Model. 

This two-photon exchange process involves long-distance contributions at the scale of low-energy QCD.  It is inherently non-perturbative but is accessible to a lattice QCD calculation~\cite{Christ:2020bzb,Chao:2024vvl,Boyle:2025fug} and it is such a calculation that is the topic of the present paper.  This calculation can be performed using a four-Fermi effective weak Hamiltonian~\cite{Buchalla:1995vs} combined with insertion of the four electromagnetic currents, needed for the emission and absorption of the two exchanged photons.  

A theoretically appealing effective theory to use in this calculation would be the four-flavor theory which describes the weak interactions at an energy scale below the bottom quark mass.  When combined with the electromagnetic interactions, this theory is renormalizable to first order in four-flavor $\Delta S=1$ weak interactions and to all orders in E\&M because of the Glashow-Iliopoulos-Maiani (GIM) cancellation.  This cancellation requires that we neglect the ratio of CKM matrix elements $|\tau| = |V_{ts}^*V_{td}/(V_{us}^*V_{ud})| = 0.00163(5)$,  which measures the failure of first- and second-column orthogonality of the CKM matrix if the top quark contribution is neglected~\cite{ParticleDataGroup:2018ovx}.

Unfortunately, a four-flavor lattice calculation of a complicated  quantity such as the long-distance, two-photon $K_L\to\mu^+\mu^-$ (LD2$\gamma$) decay amplitude is impractical at present because of the large range of scales between the charm quark and pion masses.  Thus, practical considerations require that we use instead the three-flavor effective theory of the weak interactions, valid at energies below $m_c$.  When the electromagnetic interactions are combined with the three-flavor theory the result is not renormalizable and, for the case of $K_L\to\mu^+\mu^-$ decay, additional low energy constants (LECs), whose number depends on the three-flavor regularization scheme, must be added to obtain a physical result.  These constants introduce the charm quark mass dependence which must be present in the physical result.

As we show below, this strategy solves the problem of the large range of energy scales present in the four-flavor theory. Working on a small volume with heavier than physical $u$ and $d$ quark masses we can equate quantities computed using the three- and four-flavor theories to determine the needed low energy constants.  The small volume and unphysical quark masses make the four-flavor theory calculation practical since the range of scales is reduced.   The low energy constants appear in the three-flavor theory multiplied by local operators and their effect is to make the three- and four-flavor theory agree over the full range of quark masses and lattice volumes for which the three- and four-flavor theories should describe the same physics.

In Ref.~\cite{Boyle:2025fug} we presented the first portion of this two-part calculation.  Using a lattice ensemble with a relatively coarse lattice spacing, the ``24ID'' RBC/UKQCD physical quark mass ensemble with $1/a = 1.023$ GeV, we did an essentially complete calculation of the LD2$\gamma$ decay amplitude.  However, no renormalization was performed so that instead of the necessary low energy constants, this calculation had finite but unphysical constants which depended on the lattice spacing.  In the current paper we identify these constants and describe renormalization conditions that can be imposed using both the 24ID ensemble and a second lattice ensemble with a sufficiently small lattice spacing that a charm quark can be consistently included.  These normalization conditions can be determined in less expensive calculations than those performed in Ref.~\cite{Boyle:2025fug} and will determine the terms that must be added to the results of that paper to complete the $K_L\to\mu^+\mu^-$ decay calculation.

This paper is organized as follows.  In Section~\ref{sec:renorm-overview}, we describe the counter terms that are required for a consistent three-flavor theory and the expansion in powers of the strong coupling constant $\alpha_s$ evaluated at the energy scale of the charm quark mass, identified as $\alpha_s(m_c)$ below, that can be used to simplify their calculation.  In Section~\ref{sec:Class-A}, we address the renormalization of the $\Delta S=1$ two-quark and one-photon vertex, sub-diagrams which we refer to as Class-A, and consider two possible regularization schemes.  Here the dominant sub-diagram is referred to as having a Type-1 quark contraction topology and the structure of vacuum polarization graph.  
In Section~\ref{sec:Class-B}, we discuss Class-B, $\Delta S=1$ sub-diagrams which have two quark and two photon external lines.  These appear first in our Type-3 quark contraction topology which contains a triangle graph.  This is closely related to the historical perturbative calculation of the anomalous axial Ward identity.  Section~\ref{sec:Class-C} discusses Class-C sub-diagrams which is the case where the entire diagram is divergent and a $\Delta S=1$, two-quark-two-muon counter term is needed.  Concluding remarks are made in Section~\ref{sec:conclu}.

Further technical details which are useful for the lattice calculation of these low-energy constants are given in the appendices.  In Appendix~\ref{sec:sd-kernel} we derive a more convenient analytic expression for the dispersive leptonic kernel.  In Appendix~\ref{sec:4Dto5D} we provide the necessary formulae to reconstruct a 5D propagator from a 4D, physically-projected point-source (M\"obius) Domain Wall Fermion ((M)DWF) propagator.

\section{New divergent sub-diagrams appearing in the three-flavor theory.}\label{sec:renorm-overview}

In this section we briefly review the relation between the four- and three-flavor effective theories of the weak interaction including the addition of electromagnetism (E\&M).  We then discuss the individual divergent sub-diagrams that require renormalization in the three-flavor theory when E\&M is added and their relation to the usual classification of diagrams by the topology of the quark operator contractions that is standard when describing a lattice QCD calculation.

\subsection{Three- and four-flavor weak interactions}
\label{sec:effective-weak-interactions}
As explained in detail in Section~VI of the review of Buchalla {\it et al.}~\cite{Buchalla:1995vs}, the structure of the low-energy effective theory of the non-leptonic weak interactions is determined by the small size of the $t$ and $b$ quark couplings to the $u$, $d$, $c$ and $s$ quarks, parametrized by $\tau$.  We will call the neglect of $\tau$ the approximation of Cabibbo unitarity.  In this approximation the effective $\Delta S=1$ weak interactions, which arise from $W^\pm$ exchange are described by the four current-current operators:
\begin{eqnarray}
Q_1^q = \overline{s}^i\gamma^\mu(1-\gamma^5) q_j \overline{q}^j\gamma^\mu(1-\gamma^5) d_i \\
Q_2^q = \overline{s}^i\gamma^\mu(1-\gamma^5) q_i \overline{q}^j\gamma^\mu(1-\gamma^5) d_j,
\end{eqnarray}
where $q$ equals $u$ or $c$ while $i$ and $j$ are color indices. These four operators then appear in the $\Delta S = 1$ weak effective Hamiltonian in the combination
\begin{equation}
\mathcal{H}^{\Delta S=1}(\mu) = \frac{G_F}{\sqrt{2}}\cos\theta_c\sin\theta_c \Big[z_1(Q^u_1-Q_1^c)
                     +z_2(Q^u_2-Q_2^c)\Big] \quad m_c<\mu \ll M_W.
                     \label{eq:H_W-4}
\end{equation}
Here $z_1$ and $z_2$ depend on the renormalization scale $\mu$ used to define the four operators $Q_i^q$ and $\theta_c$ is the Cabbibo angle.  Equation~\eqref{eq:H_W-4} is valid for $m_c < \mu < M_W$.  Because of their $u-c$ structure these operators do not mix with the four gluonic penguin operators whose contribution is suppressed by a factor of $\tau$.

For $\mu < m_c$ the GIM cancellation no longer applies and the four gluonic penguin operators
\begin{eqnarray}
Q_3 = \overline{s}^i\gamma^\mu(1-\gamma^5) d_i 
        \sum_{q=u,d,s}\overline{q}^j\gamma^\mu(1-\gamma^5) q_j \\
Q_4 = \overline{s}^i\gamma^\mu(1-\gamma^5) d_j 
        \sum_{q=u,d,s}\overline{q}^j\gamma^\mu(1-\gamma^5) q_i \\
Q_5 = \overline{s}^i\gamma^\mu(1-\gamma^5) d_i 
        \sum_{q=u,d,s}\overline{q}^j\gamma^\mu(1+\gamma^5) q_j \\
Q_6 = \overline{s}^i\gamma^\mu(1-\gamma^5) d_j 
        \sum_{q=u,d,s}\overline{q}^j\gamma^\mu(1+\gamma^5) q_i
\end{eqnarray}
will appear in $\mathcal{H}^{\Delta S=1}(\mu)$ at order $\alpha_s(m_c)$.  The largest Wilson coefficient is that of $Q_4$ which in Ref.~\cite{RBC:2023ynh} is found to be 2.3\% of the coefficient of $Q_2$.  Therefore, in this calculation we consider only the matrix elements of the two operators $Q_1$ and $Q_2$ and use the three-flavor effective Hamiltonian
\begin{equation}
    \mathcal{H}_{N_f=3}^{\Delta S=1} = \frac{G_F}{\sqrt{2}}\cos\theta_c\sin\theta_c \Big[z_1Q_1+z_2 Q_2\Big]     \label{eq:H_W-3}
\end{equation}

Without GIM cancellation the effective theory defined by adding $\mathcal{H}_{N_f=3}^{\Delta S=1}$ to QCD + QED is not renormalizable and counter terms must be included in $\mathcal{H}_{N_f=3}^{\Delta S=1}$ if it is to correctly describe the weak interactions to first order in the Fermi constant $G_F$.  Specifically,  there must be a separate counter term for each sub-diagram appearing in the $K_L\to\mu^+\mu^-$ calculation with a non-negative degree of divergence, including as necessary the entire diagram.  

\subsection{Divergent sub-diagrams in the three-flavor theory}
\label{subsec:divergent-diags}

\begin{figure}[h!]
\includegraphics[scale=0.5]{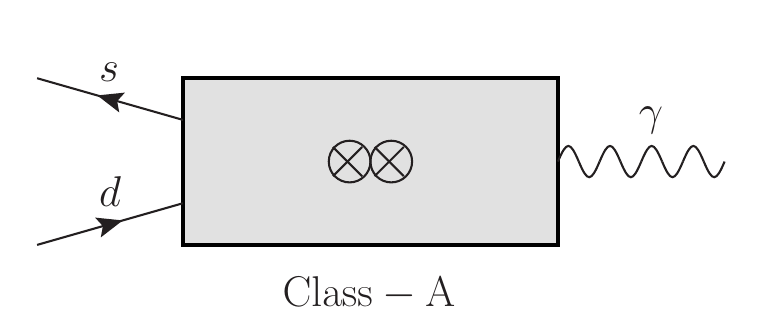} \hspace{0.5 in}
\includegraphics[scale=0.5]{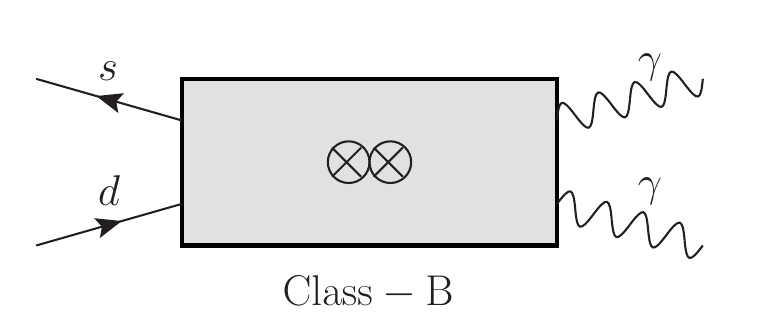} \hspace{0.5 in}
\includegraphics[scale=0.5]{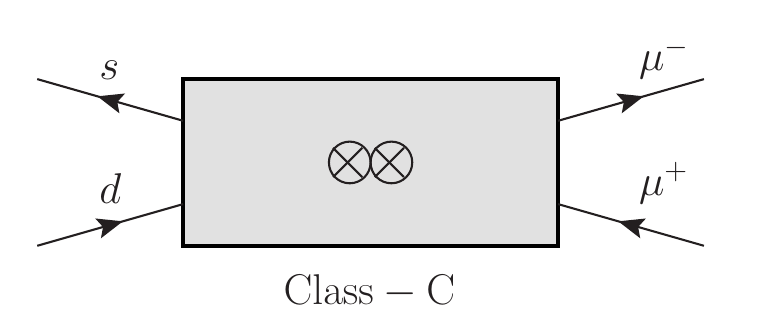}
\caption{The three classes of sub-diagram with non-negative degree of divergence which will appear in a three-flavor calculation of $K_L\to\mu^+\mu^-$ decay. The doubled cross indicates that a weak-interaction vertex is included in the sub-diagram.}
\label{fig:subdiagrams}
\end{figure}

As shown in Figure~\ref{fig:subdiagrams}, there are three types of sub-diagrams that require new counter terms to be included in $\mathcal{H}_{N_f=3}^{\Delta S=1}$.  They are determined by the following requirements.  First, any sub-diagram that requires counter terms must contain the weak vertex  corresponding to $\mathcal{H}_{N_f=3}^{\Delta S=1}$.  Second, since $\mathcal{H}_{N_f=3}^{\Delta S=1}$ changes strangeness, such a sub-diagram must have at least two external quark lines.  Third, the usual renormalization of the three-flavor $Q_1$ and $Q_2$ operators is already a standard part of the $K_L\to\mu^+\mu^-$ calculation.  Therefore, the only sub-diagrams requiring new counter terms will be those with one or two additional insertions of the electromagnetic current.  

Sub-diagrams with two external quark lines, one internal $\mathcal{H}_{N_f=3}^{\Delta S=1}$ vertex and only one external current vertex will have a degree of divergence +2 and will be labeled Class-A sub-diagrams.  
For such sub-diagrams, there are many ways to add external lines or factors of external momenta or masses which will leave the resulting sub-diagram with a non-negative degree of divergence which implies that without some simplifying step many counter terms may be required.
To reduce this complexity, we will consider two regularization schemes.  In the first scheme we will couple the external photon line to a conserved E\&M current so that an explicit factor of the external photon momentum must be extracted, reducing the degree of divergence to +1.  As worked out below, in that case three distinct counter terms may be required.  

The second scheme uses the simpler, local E\&M current -- often considered in lattice QCD -- which is not conserved. In this scheme the degree of divergence of a Class-A sub-diagram is reduced by introducing an unphysical charm quark, $\tilde{c}$, as a Pauli-Villars-type regulator in the three-flavor theory. The weak interactions of $\tilde{c}$ are identical to those of a physical charm quark to precisely enable a GIM-like ($\widetilde{\rm GIM}$) mechanism.
The restrictions on its mass $m_{\widetilde{c}}$ that must be obeyed if a simple counter term can be used to correct for the difference between $m_{\widetilde{c}}$ and $m_c$ is discussed in Section~\ref{sec:Class-A-nonconserved} below.
For a theory with chiral quarks, this $\widetilde{\mathrm{GIM}}$ subtraction will reduce the degree of divergence by two units so only a single counter term is needed to make the calculation finite.

A sub-diagram with two external quark lines, two external E\&M current vertices and an internal $\mathcal{H}_{N_f=3}^{\Delta S=1}$ vertex will have a degree of divergence +1 and is identified as Class-B.  Adding a further external gluon line will reduce the degree of divergence to -1 because of QCD gauge invariance, so that no counter term will be needed for a new class of sub-diagram with an extra gluon line. For a chiral theory, the apparent divergence for large momentum will cancel between positive and negative values of an integrand that is necessarily odd in the integration momentum. There is an ambiguity associated with the choice of integration origin needed to calculate such a linearly divergent integral~\cite{Adler:1969gk}.  As will be discussed at length in Section~\ref{sec:Class-B}, such Class-B sub-diagrams will require a single correction term. Note that, in contrast to the case of the Class-A diagrams where we have recognized scheme with different regularization for the E\&M currents, in this discussion we assume that in the lattice formulation of the calculation, a local, non-conserved electromagnetic current will be used so that adding an external photon line will reduce the degree of divergence by only one unit.

The third sub-graph topology, referred to as Class-C, that will require a new counter term is that involving the entire  $K_L\to\mu^+\mu^-$ decay amplitude.  In this case of an improper sub-diagram corresponding to the entire graph, there are four external fermion lines, two quarks and two muons, and a degree of divergence zero. Here a single counter term of the form $(\overline{s}\gamma^\mu\gamma^5 d)(\overline{\mu}\gamma^\mu\gamma^5\mu)$ is needed to make these sub-diagrams physical.  

\subsection{Quark contraction topologies and divergent sub-diagrams}

\begin{figure}[h!]
\includegraphics[scale=0.45]{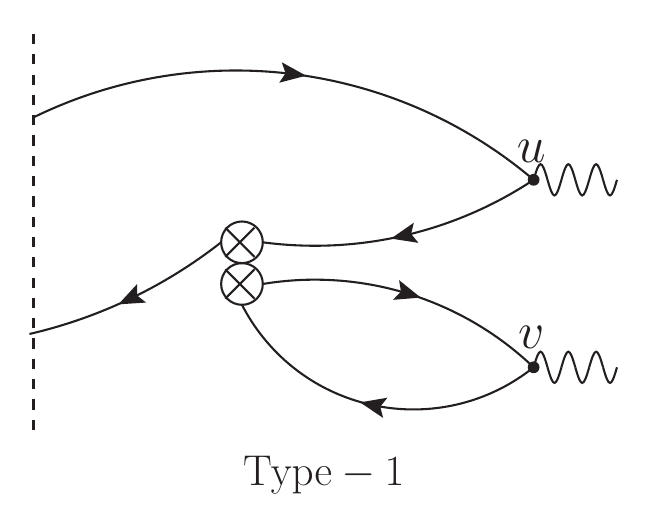} \hspace{0.2 in}
\includegraphics[scale=0.45]{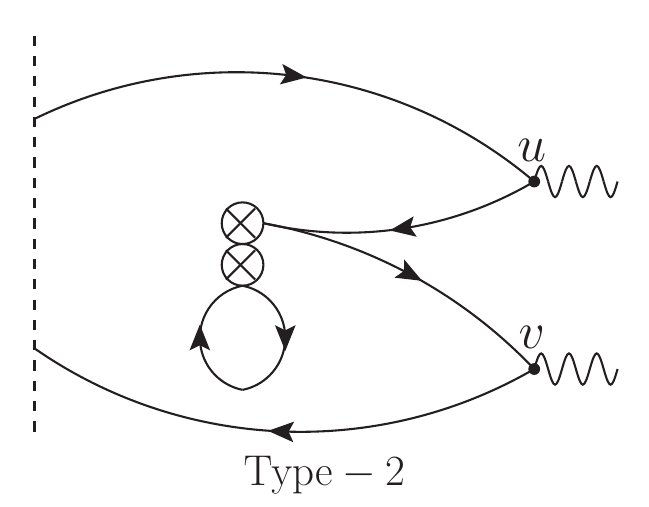} \hspace{0.2 in}
\includegraphics[scale=0.45]{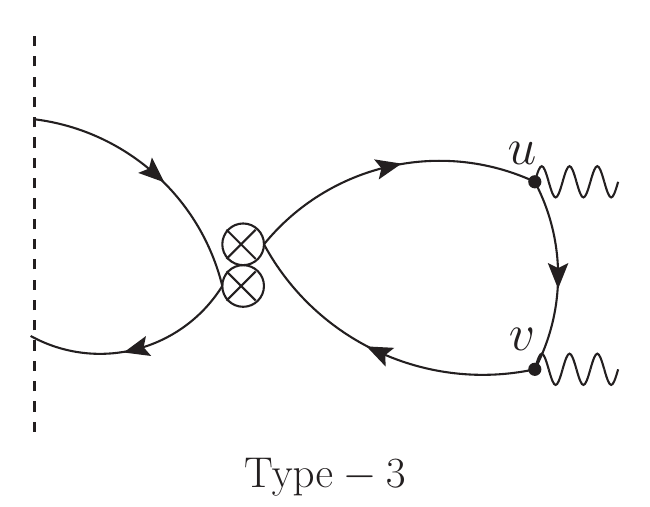} \hspace{0.2 in}
\includegraphics[scale=0.45]{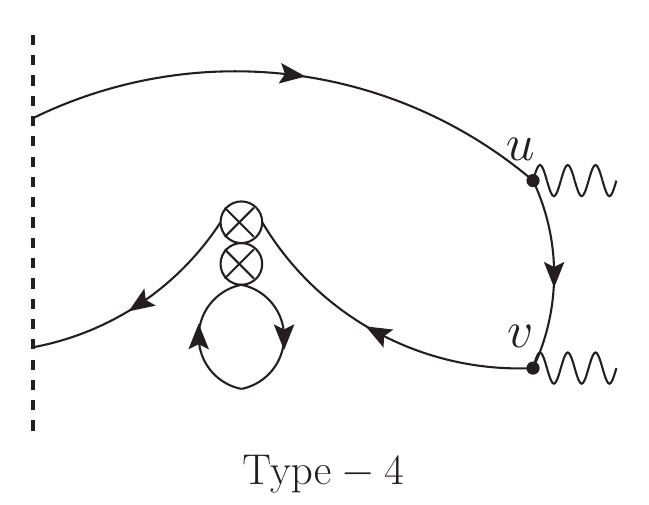} \hspace{0.2 in}
\includegraphics[scale=0.45]{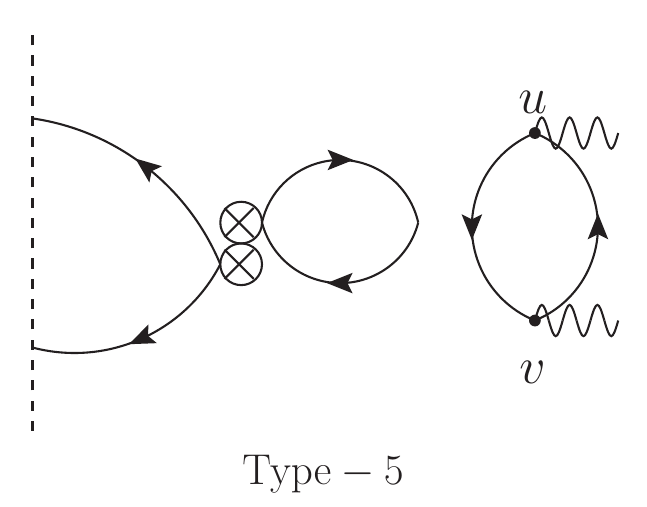} \hspace{0.2 in}
\caption{The five types of Wick-contraction topology analyzed in this paper. From left to right, top to bottom: Type-1, Type-2, Type-3, Type-4 and Type-5.  The vertical dashed line represents the Coulomb-gauge-fixed kaon interpolating operator, the pair of circled crosses the four-quark weak operator and the vertices labeled $u$ and $v$ the two E\&M currents.}
\label{fig:diagrams}
\end{figure}

In a lattice QCD calculation, the quark propagators are introduced and combined according to the usual Feynman rules while the gluon degrees of freedom, on which the quark propagators depend, are stochastically sampled using Markov chain Monte Carlo integration.  Consequently a lattice calculation of $K_L\to\mu^+\mu^-$ decay is typically described by enumerating the different topologies of quark contractions needed to construct the decay amplitude.  Figure~\ref{fig:diagrams} shows the five different quark-line topologies that appear in the lattice calculation of $K_L\to\mu^+\mu^-$ decay undertaken in Ref.~\cite{Boyle:2025fug}.

The relation between the three sub-diagrams shown in Figure~\ref{fig:subdiagrams} and the five quark-line topologies shown in Figure~\ref{fig:diagrams} is somewhat subtle since the perturbative identification of the sub-diagrams shown in Figure~\ref{fig:subdiagrams} is highly dependent on the arrangement of gluon lines which does not appear in the non-perturbative diagrams shown in Figure~\ref{fig:diagrams}.  The relation between these two descriptions of the $K_L\to\mu^+\mu^-$ decay is summarized in Table \ref{tab:sub-diag-contract}.  In that table we show the number of gluon lines which must be added to the quark propagators shown for each type of quark-line contraction appearing in Figure~\ref{fig:diagrams} to create a perturbative QCD graph which includes the sub-diagram corresponding to that column.  Since the counter terms corresponding to a particular sub-diagram are introduced to remove the unphysical, large-momentum contribution coming from that sub-diagram, each of these added gluon lines will introduce a factor of $\alpha_s(m_c)$ evaluated at the energy scale of the lattice cut-off which will typically be on the order of the charm quark mass.

\begin{table}
\centering
\begin{tabular}{l | c | c | c}
contraction\;\; & \;\;sub-diag A\;\; & \;\;sub-diag B \;\; & \;\;sub-diag C \\
\hline
Type-1      &      0     &      1     &    0       \\ 
Type-2      &      1     &      2     &    1       \\ 
Type-3      &      X     &      0     &    0       \\ 
Type-4      &      1     &      1     &    1       \\ 
Type-5      &      X     &      2     &    2 
\end{tabular}
\caption{For each of type of quark-line topology we identify which of the possible sub-diagram classes with non-negative degree of divergence can appear.  The integer shows the lowest order in $\alpha_s(m_c)$ at which that sub-diagram will appear.  An `X' indicates a sub-diagram class which does not appear for that particular quark-line topology at any order in $\alpha_s(m_c)$ .}
\label{tab:sub-diag-contract}
\end{table}

We now explain the contents of Table~\ref{tab:sub-diag-contract} by briefly considering each type of quark-line topology shown in Figure \ref{fig:diagrams} in turn.  

\begin{figure}[h!]
\includegraphics[scale=0.60]{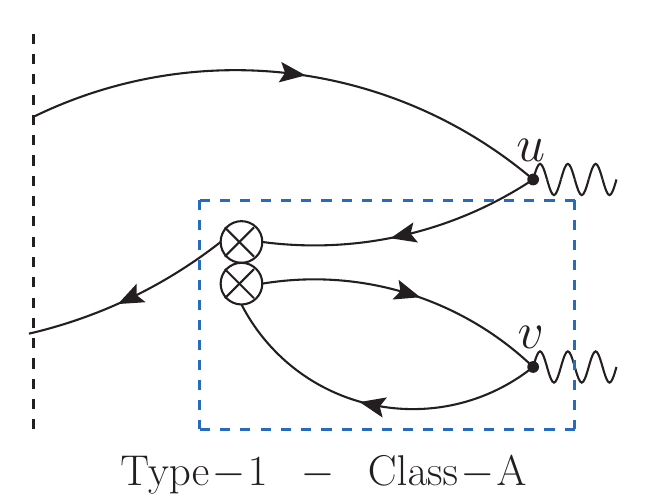} \hspace{0.75 in}
\includegraphics[scale=0.6]{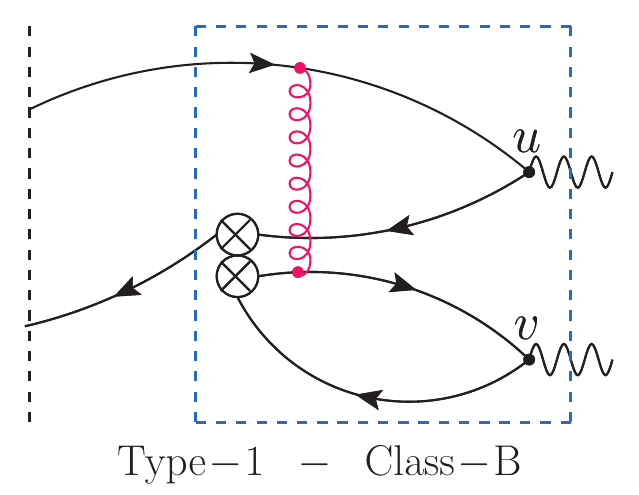} \caption{The dashed rectangles contain sub-diagrams with a non-negative degree of divergence.  The left topology has degree of divergence +2 while the right has +1.  The sub-diagram of Class-A identified on the left requires no added gluon lines and enters at order $\alpha_s(m_c)^0$.  The sub-diagram on the right is of Class-B but only after an order $\alpha_s(m_c)$ gluon line has been added.}
\label{fig:subdiagram1}
\end{figure}

\underline{\bf Type-1}\;\;In Figure~\ref{fig:subdiagram1} we identify the two classes of sub-diagram which can be found in diagrams with the Type-1 quark-line topology.  The rectangle in diagram on the left contains a sub-diagram of Class-A with degree of divergence +2.  No gluon lines need to be added to this diagram to allow this interpretation so Table~\ref{tab:sub-diag-contract} shows that a sub-diagram of Class-A appears at zeroth order of $\alpha_s(m_c)$.  As shown in the diagram on the right side of that figure, a sub-diagram of Class-B can also appear in a Type-1 diagram if a single gluon line is introduced.  This implies that if we work to first order in $\alpha_s(m_c)$ then diagrams of Type-1 will also require counter terms of Class-B. Since the entire Type-1 diagram has degree of divergence 0 and contributes directly to the $K_L\to\mu^+\mu^-$ decay without additional gluon lines, this quark-line topology includes a sub-diagram of Class-C at zeroth order in $\alpha_s(m_c)$.

\begin{figure}[h!]
\includegraphics[scale=0.6]{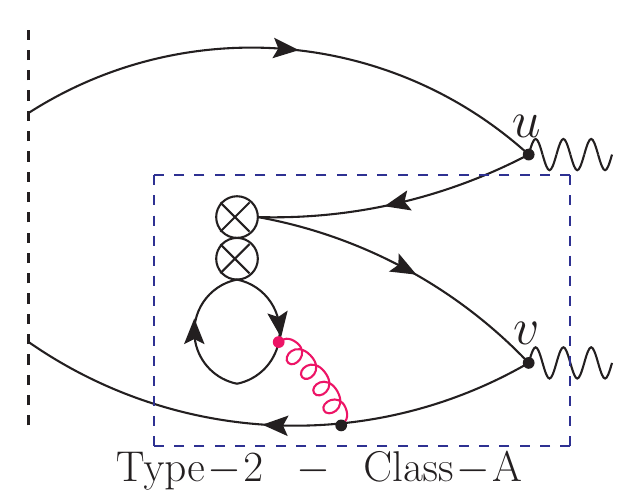} \hspace{0.5 in}
\includegraphics[scale=0.6]{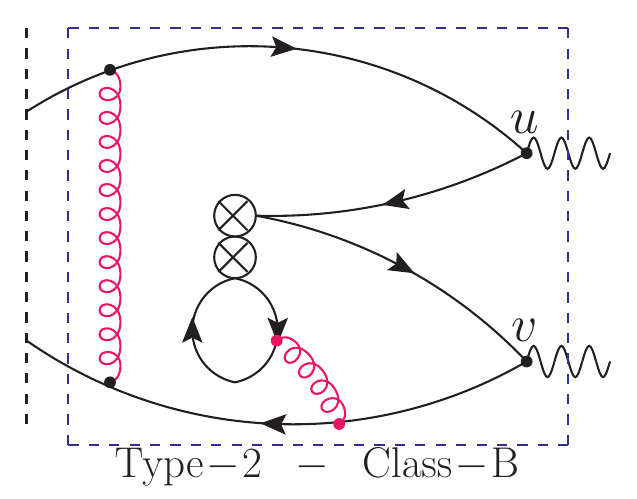} \caption{The two classes of sub-diagram with non-negative degree of divergence which can appear in graphs of the Type-2 quark-line topology to lowest order in $\alpha_s(m_c)$.  The left sub-diagram has degree of divergence +2 and appears at order $\alpha_s(m_c)$ while that at the right has degree of divergence +1 and is of order $\alpha_s(m_c)^2$.}
\label{fig:subdiagrams2}
\end{figure}

\underline{\bf Type-2}\;\; In contrast with those of Type-1, diagrams with the quark-line topology of Type-2 contain no sub-diagram with non-negative degree of divergence if no gluon lines are added.  If a gluon line is added as in the left-hand diagram in Figure~\ref{fig:subdiagrams2}, then the resulting sub-diagram contained in the dashed rectangle is Class-A with degree of divergence +2.  If a second gluon line is added as in the right portion of that figure then a sub-diagram of Class-B, enclosed in the dashed rectangle in that diagram, is created with a +1 degree of divergence.  
If no gluon line is added, then the entire graph does not contribute because the weak operator reduces to a total divergence which has no effect on the $K_L\to\mu^+\mu^-$ decay amplitude (cf. Section~III.~B.~2 of Ref.~\cite{Boyle:2025fug}).  
This implies that the entire graph will have degree of divergence 0 only at order $\alpha_s(m_c)$.

\begin{figure}[h!]
\includegraphics[scale=0.6]{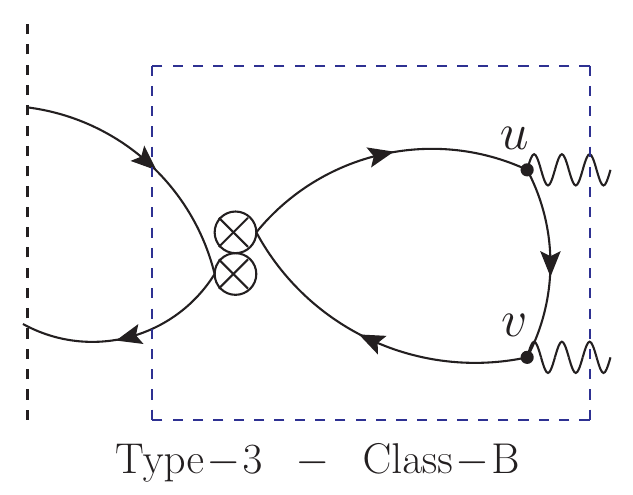} \caption{The large dashed rectangle encloses a sub-diagram of Class-B contained in diagrams of Type-3.  No additional gluon lines are required.  Likewise the entire graph, without additional gluon lines has degree of divergence 0.}
\label{fig:subdiagrams3}
\end{figure}

\underline{\bf Type-3}\;\;As shown in Figure~\ref{fig:subdiagrams3} the Type-3 quark-line topology contains a Class-B subgraph with a degree of divergence +1 without the addition of any gluon lines.  Thus, Type-3 diagrams will require a counter term at zeroth-order in $\alpha_s(m_c)$.  When the two photon and one muon internal lines are added and we consider the entire graph, this diagram will have degree of divergence zero and require a Class-C counter term, also at zeroth order in $\alpha_s(m_c)$.

\begin{figure}[h!]
\includegraphics[scale=0.6]{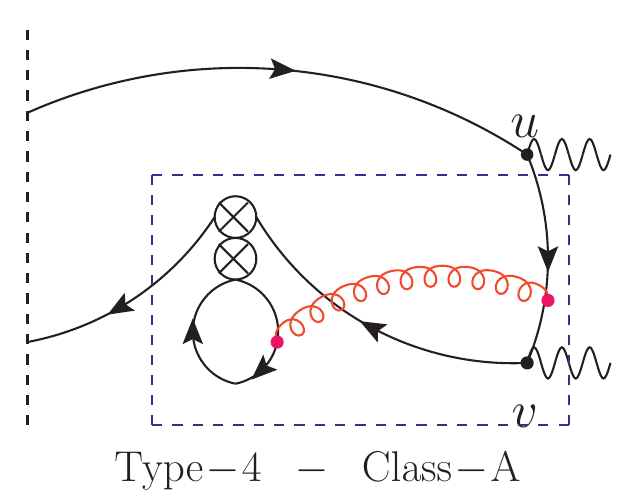} \hspace{0.5 in}
\includegraphics[scale=0.6]{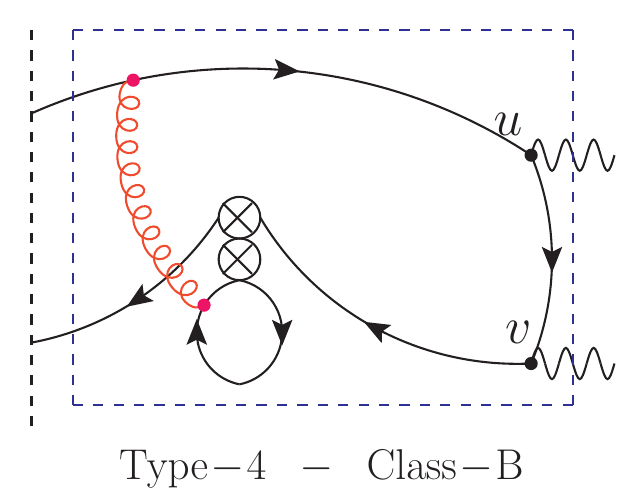} \caption{The two types of sub-diagram enclosed by dashed rectangles have non-negative degree of divergence and appear in Type-4 quark-line topologies.  The sub-diagram on the left is of Class-A and has degree of divergence +2 while the sub-diagram on the right is Class-B and has degree of divergence +1.}
\label{fig:subdiagrams4}
\end{figure}

\underline{\bf Type-4}\;\; In Figure~\ref{fig:subdiagrams4} we enclose in dashed rectangles two sub-diagrams with a non-negative degree of divergence for the quark-line topology of Type-4.  The sub-diagram identified on the left is of Class-A with degree of divergence +2 while a Class-B subdiagram with degree of divergence +1 is identified on the right.  Since each requires the addition of a single gluon line, each is of order $\alpha_s(m_c)$.  Without a gluon line the entire graph does not contribute so the required entire-graph counter term will also enter at order $\alpha_s(m_c)$.

\begin{figure}[h!]
\includegraphics[scale=0.6]{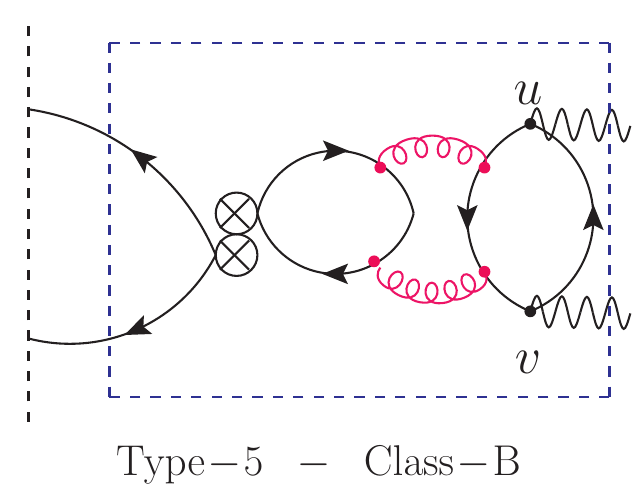} 
\caption{The dashed rectangle encloses the Class-B sub-diagram with degree of divergence +1 that appears in the Type-5 quark-line topology when two gluon lines are added to the diagram.  These are also required if the entire graph is to have non-negative degree of divergence so the counter terms needed for the Type-5 topology will all be of order $\alpha_s(m_c)^2$.}
\label{fig:subdiagrams5}
\end{figure}

\underline{\bf Type-5}\;\; Diagrams with Type-5 topology require no additional counter terms even at order $\alpha_s(m_c)$.  Sub-diagrams which do not include the quark loop joining the two E\&M currents are those of QCD + $\mathcal{H}_{N_f=3}^{\Delta S=1}$ which have already been properly renormalized by the standard treatment of the operators $Q_1$ and $Q_2$.  If a high-energy sub-graph includes the quark loop with the two E\&M currents then a minimum of two gluons must join the two parts of the graph and introduce a factor of $\alpha_s(m_c)^2$.  Figure~\ref{fig:subdiagrams5} shows the Class-B sub-diagram in this case of two-gluon exchange with its +1 degree of divergence. Likewise, the entire graph will also have a degree of divergence of 0 beginning at order $\alpha_s(m_c)^2$.

We conclude that this three-flavor calculation of $K_K\to\mu^+\mu^-$ decay requires the introduction of counter terms to compensate for three classes of sub-diagram with degrees of divergence of +2, +1 and 0.  In the next three sections we will discuss each of these three classes of sub-diagram in turn and the methods that can be used to renormalize them.

\section{Class-A sub-diagrams}
\label{sec:Class-A}

In this section we discuss the counter terms that need to be added to renormalize the Class-A sub-diagrams and how the low energy constants corresponding to each can be determined.  This section is divided into two main parts.  In the first we discuss a calculation that introduces a conserved E\&M current which reduces the degree of divergence of the Class-A sub-diagrams to +1.  In the second part we discuss an alternative strategy in which a local, non-conserved E\&M current is used and an unphysical light charm-like quark is introduced.

\subsection{Conserved E\&M current}
\label{sec:ClassA-conserved}

\subsubsection{Class-A counter terms (conserved E\&M current)}
\label{sec:ClassA-conserved-CT}

\begin{figure}[h!]
\includegraphics[scale=0.45]{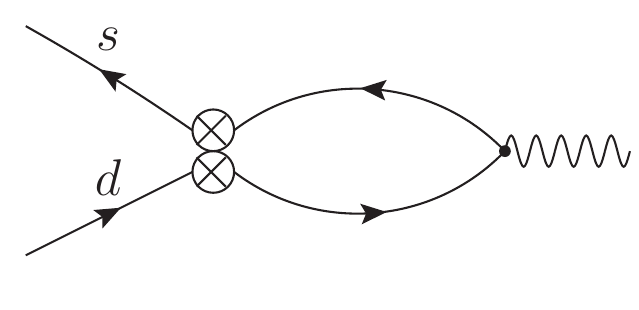} 
\hspace{0.1 in}
\includegraphics[scale=0.45]{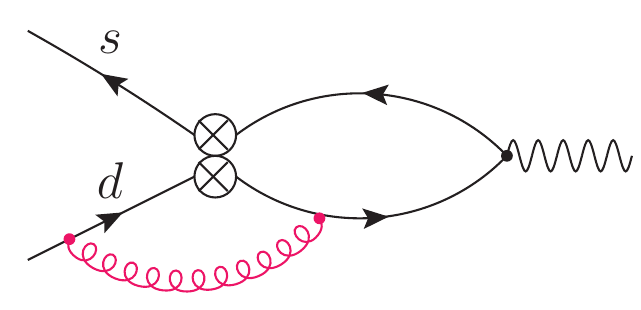}
\hspace{0.1 in}
\includegraphics[scale=0.45]{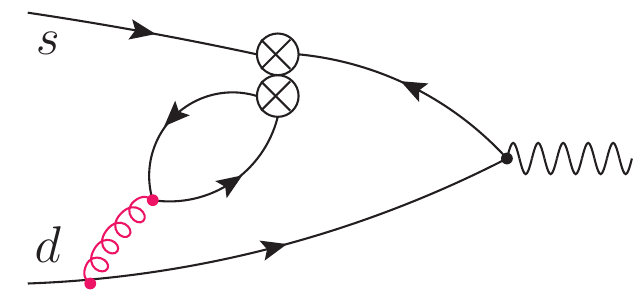}
\caption{Three examples of Class-A subgraphs with degree of divergence +2 that must be renormalized by adding counter terms whose coefficients are chosen so that the renormalized operator $\mathcal{H}_{N_f=3}^{\Delta S=1}$ will describe the physical three-flavor theory.}
\label{fig:class-A}
\end{figure}

Figure~\ref{fig:class-A} shows three examples of Class-A sub-graphs that appear at zeroth and first order in $\alpha_s(m_c)$.  The left-most two diagrams appear in Type-1 quark contractions and the right-most in Type-2.
Because $\mathcal{H}_{N_f=3}^{\Delta S=1}$ has dimension six, these three diagrams have a naive degree of divergence +2.  If the regulator used to make these diagrams well-defined is consistent with the conservation of the E\&M current then an additional external photon momentum must be extracted from each amplitude reducing their degree of divergence to +1.   

Since the implementation of the conserved current in a lattice theory of chiral fermions is significantly non-local and requires additional contact terms, one might be concerned because the term in left-hand diagram of Type-2 shown in Figure~\ref{fig:subdiagrams2} shows one of two distinct sub-diagrams of Class-A whose presence requires both currents to be conserved.  However, we can argue that the potentially troublesome contact terms that are needed to maintain current conservation when the two E\&M currents overlap  will result in a divergence only of Class-B, a divergent sub-diagram whose renormalization does not require the currents to be conserved (see Section~\ref{sec:Class-B}), so that these added contact terms can be safely omitted.

The counter terms needed to renormalize the Class-A sub-diagrams must be hermitian, conserve parity and charge conjugation, respect chiral symmetry and be symmetrical under the exchange of $d$ and $s$ quarks.  These requirements imply that only three counter terms are possible:
\begin{eqnarray}
    Q_1^A &=& \left(\overline{s}\gamma^\nu d + \overline{d}\gamma^\nu s\right) \partial^\mu F^{\mu\nu} \label{eq:class-A-1} \\
    Q_2^A &=&  \frac{1}{2}\left\{\overline{s}\gamma^\nu\gamma^5(\overleftarrow{D}^\mu - \overrightarrow {D}^\mu) d
    + \overline{d}\gamma^\nu\gamma^5(\overleftarrow{D}^\mu - \overrightarrow {D}^\mu) s\right\}\, \widetilde{F}^{\mu\nu} \label{eq:class-A-2} \\
    Q_3^A &=& (m_d+m_s)\left(\overline{s}\sigma^{\mu\nu} d +\overline{d}\sigma^{\mu\nu} s\right) F^{\mu\nu}, \label{eq:class-A-3}
\end{eqnarray}
where $\widetilde{F}^{\mu\nu}$ is the dual of $F^{\mu\nu}$.  The operator $Q_1^A$ will take the form of an electroweak penguin operator if the equations of motion are used to replace the right-most factor, $\partial^\mu F^{\mu\nu}$, in Eq.~\eqref{eq:class-A-1} by the  electromagnetic current resulting from the quark degrees of freedom~\cite{DAmbrosio:2018ytt}. 
In the present context, we keep only the electromagnetic current of the muon that also appears when $\partial^\mu F^{\mu\nu}$ is evaluated.  

The sub-diagrams of Class-A can then be made consistent with a physical four-flavor theory by including the combination
\begin{equation}
\sum_{i=1}^3 C^A_i O^A_i
\end{equation}
to the unrenormalized operator $\mathcal{H}_{N_f=3,\mathrm{lat}}^{\Delta S=1}$ with an appropriate choice of the three coefficients $C^A_i$.  Here we have added the label ``lat'' to $\mathcal{H}_{N_f=3,\mathrm{lat}}^{\Delta S=1}$ to emphasize that without these three counter terms, $\mathcal{H}_{N_f=3,\mathrm{lat}}^{\Delta S=1}$ is a partially unrenormalized quantity whose matrix elements will depend logarithmically on the lattice spacing $a$.

\subsubsection{Class-A renormalization coefficients (conserved E\&M current)}

Since to a good approximation the four-flavor theory of $K_L\to\mu^+\mu^-$ is well defined without additional renormalization constants (beyond those needed to renormalize the four-quark operators which appear in $\mathcal{H}_{N_f=4}^{\Delta S=1}$), we can determine the coefficients $C^A_i$ by comparison with these well-defined, four-flavor predictions.  
As discussed above, we can determine the three coefficients $C^A_i$ from a four-flavor calculation performed under unphysical conditions on a comparison lattice ensemble, $\mathcal{E}_{N_f=4}$. This ensemble must have a sufficiently small lattice spacing to allow the accurate treatment of a charm quark with physical mass. However, the ensemble $\mathcal{E}_{N_f=4}$ can involve heavier than physical light quarks and a correspondingly smaller physical volume.

We therefore should work with two ensembles, one with three-flavors, $\mathcal{E}_{N_f=3}$, and one with four $\mathcal{E}_{N_f=4}$.   These need not have identical lattice spacings but both lattice spacings should be sufficiently small that the effects of non-zero lattice spacing can be ignored.  Since the three- and four-flavor discretization errors will be different, if these errors are important, then two sequences of ensembles will be required and the three- to four-flavor matching done by comparing the three- and four-flavor continuum limits.  

Since the three-flavor counter terms $C_i^\mathrm{lat}Q_i^{A,\rm lat}$ define an effective three-flavor theory which agrees with the four-flavor theory at low-energy, this agreement will hold for all variants of the three- and four-flavor theories with matching low energy properties. Thus, a comparison between identical three- and four-flavor low-energy theories with unphysical light quark masses (but masses which agree between the two theories when expressed in physical units)  can be used to determine the $C_i^\mathrm{lat}$.  Since these three counter terms are introduced to renormalize the two-quark and one E\&M current, Class-A sub-diagrams such as those shown in Figure~\ref{fig:class-A}, their coefficients can be determined by matching general amplitudes with these three external lines. 
In the following we will drop the label `lat' for notational simplicity.

Given the presence of external quark lines, this comparison can be done by working with Green's functions constructed using RI/MOM~\cite{Martinelli:1994ty} renormalized quark operators and the conserved E\&M current and evaluating three- and four-flavor Landau-gauge-fixed Green's functions of the form:
\begin{equation}
    \bar{\Lambda}_{\beta\alpha,\mu}^{N_f}(p_1,p_2)
    = \int\int\int d^4x\, d^4 y\, d^4z\, e^{-ip_1y} e^{ip_2z} \left\langle s_\alpha(y) \mathcal{H}^{\Delta S=1}_{N_f}(x)J_\mu(0) \overline{d}_\beta(z) \right\rangle^{\mathrm{amp}},
    \label{eq:class_A-amplitude}
\end{equation}
where the RI/MOM quark operators $\overline{d}$ and $s$ have been multiplicatively renormalized so that the corresponding off-shell $s$ and $d$ quark propagators take a specified value at the renormalization scale. The current $J_\mu$ must be the non-local, conserved E\&M current as required by our renormalization strategy.  The superscript `amp' indicates that the usual amputation of the two external quark propagators has been performed.  The angle brackets in Eq.~\eqref{eq:class_A-amplitude} indicate an amplitude evaluated to all orders in QCD. 

In four flavors, Eq.~\eqref{eq:class_A-amplitude} is a finite quantity thanks to the GIM mechanism, which allows us to define the physical target Green's function
\begin{eqnarray}
    \Lambda_{\beta\alpha,\mu}^{\nf = 4}(p_1,p_2) \equiv \bar{\Lambda}_{\beta\alpha,\mu}^{\nf = 4}(p_1,p_2)\,.
\end{eqnarray}
On the contrary, additional counter terms Eqs.~(\ref{eq:class-A-1} - \ref{eq:class-A-3}) are needed to eliminate the short-distance divergence and restore the correct physical contribution in the three-flavor case, with the renormalized Green's function being defined as 
\begin{eqnarray}
\Lambda_{\beta\alpha,\mu}^{N_f=3}(p_1,p_2) &\equiv& 
\bar{\Lambda}_{\beta\alpha,\mu}^{N_f=3}(p_1,p_2) + 
\sum_{i=1}^3 C_{i}^{A}\, \int\int\int d^4 xd^4 y\, d^4z\, e^{-ip_1y} e^{ip_2z} e^{ix(p_1-p_2)} \label{eq:determine_C_i}\\
    && \hskip 1.8 in \times \left\langle s_\alpha(y)Q_i^{A}(0)
    \overline{d}_\beta(z) A_\mu(x) \right\rangle^{\mathrm{amp}}\,,
    \nonumber
\end{eqnarray}
where the Green's function in the last line has the two external quark and photon legs amputated.

In a three-flavor calculation accurate to arbitrary order in $\alpha(m_c)$ we must evaluate the Green's function given in Eq.~\eqref{eq:class_A-amplitude} using all quark-line topologies which include a Class-A sub-diagram.  
We would evaluate this Green's function with unphysically large light quark masses on the three- and four-flavor ensembles $\mathcal{E}_{N_f=3}$ and $\mathcal{E}_{N_f=4}$ and require equality for three independent spin and momentum combinations to determine the three $C^{\mathrm{lat}}_i$ coefficients.  

If we intend to work only to zeroth order in $\alpha_s(m_c)$ our task is less complicated.  First, since only the Type-1 quark-line topology contains sub-diagrams with a non-negative degree of divergence at order $\alpha_s(m_c)^0$, only that topology needs to be evaluated as shown for example, on the left-hand side of Figure~\ref{fig:subdiagram1}.  Second, of the three counter terms shown in Eqs.~\eqref{eq:class-A-1}, \eqref{eq:class-A-2} and \eqref{eq:class-A-3}, those in Eqs.~\eqref{eq:class-A-2} and \eqref{eq:class-A-3} contain external momenta or masses that can appear in a divergent loop only if a high-momentum gluon line attaches to an external quark propagator which introduces at least one power of $\alpha_s(m_c)$.  We can then determine the only relevant coefficient $C_1^{\mathrm{lat}}$ by imposing a single condition connecting the three- and four-flavor theories such as:
\begin{equation}
\left(\gamma^\mu\right)_{\alpha\beta}\Lambda^{\mathrm{N_f=3}}_{\alpha\beta\mu}(p_2,p_1)
=\left(\gamma^\mu\right)_{\alpha\beta}\Lambda^{\mathrm{N_f=4}}_{\alpha\beta\mu}(p_2,p_1). \label{eq:Class-A-Match} 
\end{equation}
The left-hand, three-flavor side of Eq.~\eqref{eq:Class-A-Match} contains the sum of $\mathcal{H}^{\Delta S=1}_{N_f=3,\mathrm{lat}}$ and the single counter term with the coefficient $C_1$.  This constant is then determined by the four-flavor, right-hand side which contains no unknowns.  Both sides of Eq.~\eqref{eq:Class-A-Match} must be evaluated at the same, well-defined low-energy point.  The momenta $p_i$ on the right and left must be the same in physical units.  The space-time volumes must be the same or sufficiently large that finite volume effects can be neglected.  The heavier-than-physical light quark masses must be the same in physical units,  {\it etc}.  Of course, the lattice spacing used on the left and right need not be the same if discretization errors can be ignored.

In practice, some of these requirements will be ignored, introducing systematic errors that must be estimated.  When completing the calculation of Ref.~\cite{Boyle:2025fug} the left-hand side of Eq.~\eqref{eq:Class-A-Match} would naturally be evaluated using the same 24ID domain wall fermion ensemble as used in that paper.  We plan to evaluate the right-hand side using the RBC/UKQCD 32IF ensemble~\cite{RBC:2014ntl} which has an inverse lattice spacing of 1/$a$=3.148 GeV and a light quark mass that results in a pion mass of 371 MeV.  This lattice spacing is sufficiently small to support a physical domain wall quark with the charm quark mass.  While we will use a light quark mass when computing the left-hand side of Eq.~\eqref{eq:Class-A-Match} that also results in a 371 MeV pion, there will be errors in our application of that equation.  The sea light quark mass for the 24ID ensemble corresponds to a physical mass pion so our calculation of the left-hand side will be partially quenched with the wrong light sea quark mass.  For the right-hand side the calculation will be unitary in the sense that the sea and valence light quark masses will be the same.  However, the effects of the charm quark determinant will be absent since this too is a three-flavor ensemble.

Since the possibility of working to a fixed order in $\alpha_s(m_c)$ is being explored, we should also consider using QCD pertubation thoery to evaluate the renormalization constants $C_i^A$.  This is particularly accessible for the case of an O$(\alpha_s(m_c)^0)$ calculation.  Without additional gluon lines, the Type-1 counter term could be computed from the one-loop diagram shown on the left-hand side of Figure~\ref{fig:class-A}: a simple vacuum polarization diagram.  

Of course, this should be done using regularization-independent (RI) methods to avoid the poor convergence~\cite{Dashen:1980vm} of lattice perturbation theory.  Employing an RI scheme to determine the coefficient $C_1^A$ requires that $C_1^A$ be computed as two separate terms:
\begin{equation}
    C_1^A = C_1^{A,\rm NP} + C_1^{A,\rm P}.
\end{equation}
The first, non-perturbative coefficient is determined by requiring that when the counter term $C_1^{A,\rm NP}Q_1^A$ is added to the three-flavor weak Hamiltonian in Eq.~\eqref{eq:Class-A-Match}, the resulting amplitude $\Lambda_{\alpha\beta,\mu}^{N_f=3}(p_1,p_2)$ defined in Eq.~\eqref{eq:class_A-amplitude} obeys the condition:
\begin{equation}
\left(\gamma^\mu\right)_{\alpha\beta}\Lambda^{\mathrm{N_f=3}}_{\alpha\beta\mu}(p_2,p_1) = 0 \label{eq:Class-A-perturbative}
\end{equation}
for specific values of the four-momenta $p_1$ and $p_2$.  (Here the choice that this renormalized amplitude vanish is arbitrary but convenient.)  With this counter term, we now have a three-flavor theory which obeys the RI renormalization condition given in Eq.~\eqref{eq:Class-A-perturbative}.  This is a well-defined theory for which one could evaluate a continuum limit but one with an incorrect value of the $C_1^A$ LEC.

We can then carry out a continuum, one-loop, four-flavor perturbation theory calculation of the well-defined four-flavor quantity $\left(\gamma^\mu\right)_{\alpha\beta}\Lambda^{\mathrm{N_f=4}}_{\alpha\beta\mu}(p_2,p_1)$ to determine what value the correct three-flavor theory should give for this quantity. The second, perturbative coefficient $C_1^{A,\rm P}$ would then be chosen so that when $\mathcal{H}^{\Delta S = 1}_{N_f=3}$ is replaced by the counter term $C_1^{A,\rm P}Q_1^A$ in Eq.~\eqref{eq:Class-A-Match}, one obtains the correct, perturbative four-flavor value, thus determining the value that should be used for $C_1^{A,\rm P}$ in the three-flavor theory.  Of course, for this perturbative QCD calculation of $C_1^{A,\rm P}$ the four-momenta $p_1$ and $p_2$ must sufficiently large that a perturbative calculation of the QCD amplitude $\Lambda(p_1,p_2)^{N_f=4}$ is accurate.  However, they must also lie sufficiently far below the charm quark mass that the effective three-flavor calculation is consistent and sufficiently far below the $N_f=3$ inverse lattice spacing that large discretization errors are avoided.

A similar perturbative determination of the needed counter terms is possible for all of the cases considered below except those associated with Class-B diagrams.  We view the case just discussed as a sufficiently complete example that this discussion need not be repeated for these other situations for which a perturbation theory calculation is possible.

\subsection{Local E\&M current with unphysical light charm quark, $\widetilde{c}$} \label{sec:Class-A-nonconserved}

We now discuss the second strategy to reduce the degree of divergence of Class-A sub-diagrams: using a lighter-than-physical charm quark $\widetilde{c}$ and the simpler local E\&M current.  

\subsubsection{Class-A counter terms (unphysical light charm quark)}

Here we reduce the computational complexity of Class-A sub-diagrams by using the local, non-conserved E\&M current.  The +2 degree of divergence of these diagrams is reduced by introducing an unphysical light charm quark $\widetilde{c}$ with weak interaction couplings identical to those of the physical charm quark but whose mass $m_{\widetilde{c}}$ is sufficiently small that the $\widetilde{c}$ quark can be accurately treated on the three-flavor $\mathcal{E}_{N_f=3}$ ensemble. We must also require that $m_{\widetilde{c}}$ is large compared to the scale of the $K_L\to\mu^+\mu^-$ decay so that, like the charm quark, its effects can be viewed as short-distance and represented by local counter terms in the three-flavor theory.

\begin{figure}[h!]
\includegraphics[scale=0.6]{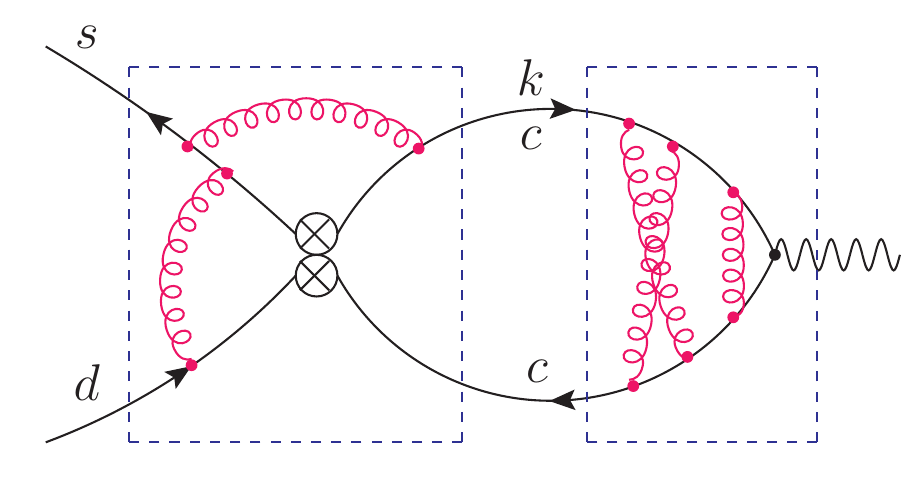}
\caption{Identification of the two components of a general Class-A sub-diagram.  While a particular graph is shown, the two dashed rectangles identify two parts of a general Class-A sub-diagram.  The rectangle on the left is required to be two-particle irreducible while that on the right simply encloses the remainder of the diagram.}
\label{fig:Class-A_loop}
\end{figure}

To determine the physical $K_L\to\mu^+\mu^-$ amplitude using this approach we must remove lattice artifacts in the three-flavor theory introduced by the use of a non-conserved current and we must determine a correction term to compensate for our use of a $\widetilde{c}$ instead of a $c$ quark. To do this we will separate a general Class-A sub-diagram into two disjoint sub-diagrams joined by two $c$ or $\widetilde{c}$ propagators.  This division is shown in Figure~\ref{fig:Class-A_loop} for a particular example.  The sub-diagram on the left is required to be two-particle irreducible with respect to the charm quark lines: the left rectangle is drawn so that the diagram it contains cannot be divided into two disjoint pieced by cutting two $c$ propagators.  The rest of the diagram is contained in the rectangle on the right.  The two charm quark lines joining these two sub-diagrams are uniquely determined and the momentum of the quark moving to the right is labeled $k$. 
The right sub-diagram contains all of the graphs making up the electromagnetic current vertex.

We use the variable $k$ to divide the momentum integration in the Class-A subdiagram into two regions.  In Region I, the magnitude of $k$ lies in the range $0 \le k \le \kappa/a$.  While in Region-II this magnitude obeys $\kappa/a \le k \le 4/a$.  
Here the upper bound $4/a$ is chosen to represent the largest allowed lattice momentum with components: $k_\mu = 2\sin(\pi/2)/a$.  The variable $\kappa$ is a small number which bounds the discretization errors in Region I which will be $(ka)^2 \le \kappa^2$.  These two momentum integration regions will be represented by different counter terms.

For Region I, the electromagnetic current vertex has bounded external momenta and will differ from the conserved current by terms of order $(ka)^2$.  We will choose $\kappa$ sufficiently small that these can be neglected. Thus, the contribution of Class-A sub-diagrams evaluated in Region-I will have the same properties as those same diagrams in the case of a conserved E\&M current and can be accurately represented at long distances, up to terms of order $\kappa^2$ by sum of the three local operators $\{O_i\}_{1 \le i \le 3}$ given in \Cref{eq:class-A-1,eq:class-A-2,eq:class-A-3}.  

Region II must be treated differently since at momenta larger than $\kappa/a$, the E\&M current vertex appearing in the right rectangle of Figure~\ref{fig:Class-A_loop} will no longer be conserved.  However, we will assume that $m_{\widetilde{c}}$ obeys the strong inequality $m_{\widetilde{c}} \ll \kappa/a$, so that in Region-II we can expand in $(m_u/k)^{2n}$ and $(m_{\widetilde{c}}/k)^{2n}$.  Because of the $\widetilde{\rm GIM}$-subtraction, the leading, constant term in this expansion in $(m_{\widetilde{c}}/k)^{2n}$ cancels and a factor of $m_u^2-m_{\widetilde{c}}^2$ remains.

After this factor of $m_u^2-m_{\widetilde{c}}^2$ is extracted the remaining amplitude has degree of divergence 0 but need not depend on the E\&M field in a gauge-invariant way.  These conditions are fulfilled by the single local operator:
\begin{equation}
O_4 = (\overline{s}\gamma^\mu d) A_\mu
\label{eq:GIM-tilde-1}
\end{equation}
Note, we can also conclude directly, at least in a chiral theory, that in the case of a GIM-like subtraction a counter term of the form given in Eq.~\eqref{eq:GIM-tilde-1} must have a coefficient that is linear in the squares of the quark masses. Two derivatives with respect to a squared quark mass would make the amplitude convergent which would imply that such a gauge-non-invariant term must have been removed. 

This use of the simpler non-conserved current and an unphysical light charm quark in the three-flavor calculation of $K_L\to\mu^+\mu^-$ decay then requires two corrections.  The first involves the three operators $\{O_i\}_{1\le i \le 3}$ which can be interpreted as describing the long-distance effects of replacing the unphysical $\widetilde{c}$ quark in  $\widetilde{\mathrm{GIM}}$ subtraction with the correct, $c$ quark.  Second we must correct for the high-momentum portion of the $u-\widetilde{c}$ loop arising from our use of a non-conserved E\&M current -- an effect which is suppressed by the $(m_u^2 - m_{\widetilde{c}}^2)$ difference.

\subsubsection{Class-A renormalization coefficients (unphysical light charm quark)}

The finite coefficient $C_4^A$ of the counter term shown in Eq.~\eqref{eq:GIM-tilde-1} which should be added to each Class-A sub-diagram which appears in the three-flavor calculation is straight-forward to determine.  We need only adjust $C_4^A$ so that after that counter term has been added, the E\&M current is conserved when the Class-A sub-diagram being computed is evaluated at energies well below the lattice scale.

Following Eq.~\eqref{eq:class_A-amplitude} we can define a new Landau gauge-fixed amplitude:
\begin{equation}
    \Lambda_{\beta\alpha,\mu}^q(p_1,p_2)
    = \int\int\int d^4x\, d^4 y\, d^4z\, e^{-ip_1y} e^{ip_2z} \left\langle s_\alpha(y) \mathcal{H}^{\Delta S=1}_q(x)J_\mu(0) \overline{d}_\beta(z) \right\rangle^{\mathrm{amp}},
    \label{eq:class_A-amplitude-q}
\end{equation}
where $\mathcal{H}^{\Delta S=1}_q(x)$ is the weak Hamiltonian in which the single up-type quark $q$ is included and, in the Class-A amplitude described by the right-hand side of this equation, that quark $q$ is contracted with itself.  Thus, this first renormalization of the Class-A amplitude being computed on the 3-flavor ensemble $\mathcal{E}^{N_f=3}$ would take the form:
\begin{eqnarray}
    \widetilde{\Lambda}_{\beta\alpha,\mu}^{N_f=3}(p_1,p_2) &=& \Lambda_{\beta\alpha,\mu}^u(p_1,p_2) - \Lambda_{\beta\alpha,\mu}^{\widetilde{c}}(p_1,p_2) 
    \label{eq:class_A-amplitude_u-c_tilde} \\ 
    && +\;C_4^A\, m_{\widetilde{c}}^2 \int\int d^4 y\, d^4z\, e^{-ip_1y} e^{ip_2z} \left\langle s_\alpha(y)(\overline{s}\gamma^\mu d)(0)
    \overline{d}_\beta(z) \right\rangle^{\mathrm{amp}}.
    \nonumber
\end{eqnarray}
The coefficient $C_4^A$ should be adjusted so that
\begin{equation}
   (p_1-p_2)^\mu \widetilde{\Lambda}_{\beta\alpha,\mu}^{N_f=3}(p_1,p_2) =0,
\end{equation}
where the index $\mu$ is summed. This equation should hold for all values of the spinor indices $\alpha$ and $\beta$ if $C_4^A$ is chosen to make it true for one pair of values, provided $p_1$ and $p_2$ lie well below $1/a$.

Next, we can correct for the use of the unphysical charm quark mass in the three-flavor calculation by including a set of three counter terms that can be non-perturbatively determined on the four-quark ensemble $\mathcal{E}_{N_f=4}$.  If we wish to correct a subset of Class-A sub-diagrams we would compute those sub-diagrams on the ensemble $\mathcal{E}_{N_f=4}$ but use a GIM-subtracted combination of $\widetilde{c}$ and the physical charm quark $c$ in the closed quark loop appearing in those diagrams.  Thus, on ensemble $\mathcal{E}_{N_f=4}$ we would compute 
\begin{eqnarray}
    \widetilde{\Lambda}_{\beta\alpha,\mu}^{N_f=4}(p_1,p_2) &=& \Lambda_{\beta\alpha,\mu}^{\widetilde{c}}(p_1,p_2) - \Lambda_{\beta\alpha,\mu}^{c}(p_1,p_2) 
    \label{eq:class_A-amplitude_c_tilde-c}.
    \nonumber
\end{eqnarray}
As discussed in the previous section this GIM-subtracted combination should be well represented by a local operator which must have the form:
\begin{equation}
\sum_{i=1}^3 C_{4+i}^A Q_i^A.
\label{eq:ctilde-c_CT}
\end{equation}

To determine the values of the three coefficients $\{C_i\}_{5 \le i \le 7}$ needed to make the counter term in Eq.~\eqref{eq:ctilde-c_CT} reproduce the effects of the $\widetilde{c}$ - $c$ subtracted weak Hamiltonian at low energies (much smaller than $m_{\widetilde{c}} < m_c$)
we need to equate the matrix element of the operator given in Eq.~\eqref{eq:ctilde-c_CT} with the difference of Class-A subdiagrams which is is supposed to represent:
\begin{eqnarray}
    \widetilde{\Lambda}_{\beta\alpha,\mu}^{N_f=4}(p_1,p_2) &=& \sum_{i=1}^3 C_{4+i}^A\, m_{\widetilde{c}}^2 \int\int\int d^4 xd^4 y\, d^4z\, e^{-ip_1y} e^{ip_2z} e^{ix(p_1-p_2)} \label{eq:determine_C_i}\\
    && \hskip 1.8 in \times \left\langle s_\alpha(y)(Q_i)(0)
    \overline{d}_\beta(z) A_\mu(x) \right\rangle^{\mathrm{amp}},
    \nonumber
\end{eqnarray}
where here the label ``amp'' indicates amputating both the two external quark propagators and the photon propagator.  This equation will determine the three coefficients $\{C_i\}_{5 \le i \le 7}$ if its left- and right-hand sides are evaluated for three independent choices of the spinor indices and external momenta and then the resulting three equations solved for the $C_i$. 

As in the previous section the operators $\{Q_i\}_{1 \le i \le 3}$ should be normalized in an RI/MOM scheme so they have a physical meaning. 
When multiplied by the coefficients $\{C_i\}_{5 \le i \le 7}$ determined from the above equation these become  well-defined physical operators whose matrix elements express the difference between performing a GIM subtraction using the heavy quark $\widetilde{c}$ and the heavier physical charm quark $c$.  These physical operators can then be included as correction terms in the three-flavor calculation provided the operators $Q^A_i$ are normalized in the same way on the three-flavor ensemble and the same coefficients $C^A_i$ are used that were determined on the four-flavor ensemble.

If we choose to use a conserved E\&M current in the four-flavor theory when determining the $\{C_i\}_{5 \le i \le 7}$, the above discussion is complete.  However, if desired, we can instead use a non-conserved current and the method just described will work as well.  However, before imposing Eq.~\eqref{eq:determine_C_i} it will be necessary to make the same correction described in Eq.~\eqref{eq:class_A-amplitude_u-c_tilde}.  We must introduce a single counter term to correct for the non-conservation of that E\&M current at the lattice scale.

The use of a non-conserved E\&M current, that this unphysical $\widetilde{\mathrm{GIM}}$ subtraction allows, has important computational advantages since the RBC/UKQCD collaboration now stores precomputed propagators in a sparsened format in which the propagator values at only a fraction of the possible sink sites are stored.  Such propagators can be readily used to evaluate contractions with a local current where only propagator values at a single lattice site are needed.  For a non-local, conserved current the propagator values will typically not be stored at the non-local collection of nearby points that are needed.  However, the requirement that an unphysical charm quark mass $m_{\widetilde{c}}$ can be found that is both smaller than the inverse lattice spacing at which the three flavor calculation is performed and yet sufficiently massive that its interactions appear local on the scale of the $K_L\to\mu^+\mu^-$ decay may be difficult to meet.  

\section{Class-B sub-diagrams}
\label{sec:Class-B}

This second class of divergent sub-diagrams, shown on the right-hand side of the top line of Figure~\ref{fig:subdiagrams}, has external strange and down quark lines and two E\&M currents coupling to external photons.  Such Class-B sub-diagrams have degree of divergence +1.  The simplest example of such a sub-diagram is the diagram formed from the Type-3 quark-line topology shown in Figure~\ref{fig:subdiagrams3} with no additional gluon lines.  Since the initial $K_L$ state and final $\mu^+\mu^-$ states are charge conjugation even and parity odd, the term in the weak interaction vertex that will contribute in this simplest case of no additional gluon lines is the product of two axial currents:
\begin{equation}
    \frac{G_F}{\sqrt{2}}(\overline{s}\gamma^\nu\gamma^5 d) (\overline{u}\gamma^\nu\gamma^5u).
\end{equation}
In this case, the left-most vertex in the up quark loop shown in Figure~\ref{fig:subdiagrams3} is the axial vector current and this simplest, one-loop Class-B subgraph is the anomalous triangle diagram analyzed by Adler~\cite{Adler:1969gk} and by Bell and Jackiw~\cite{Bell:1969ts} multiplied by an extra $\overline{s}\gamma^\mu\gamma^5 d$ vertex.  In the four-flavor theory, the anomalous behavior of this triangle diagram (and that of other higher-order graphs that appear when gluons lines are added) is eliminated by the usual GIM cancellation.  

This class of sub-diagram would also offer no special problem were we to use the conserved, non-local E\&M current in our lattice calculation.  In that case two additional powers of the external photon momenta would need to be extracted from the integrand to construct two factors of the E\&M field strength tensor $F^{\mu\nu}$.  This would reduce the degree of divergences of Class-B sub-diagrams to -1 and no further discussion would be needed.  However, the conserved vector current for domain wall fermions is non-local and involves additional contact terms~\cite{Boyle:2015vda} that have not been worked out when the product of two such currents is to be computed.  As a result for all of the quark line topologies except Type-1, we work with the non-conserved local current and must deal with a lattice regulator scheme which violates current conservation.  While not actually divergent, for a three-flavor theory without current conservation, these Class-B sub-diagrams will typically require the addition of a local correction term if a physical result is to be obtained.  We discuss this correction term in Section~\ref{sec:AdlerAmbiguity} and a method that can be used to determine it in Section~\ref{sec:ambiguity-fix}.

\subsection{Adler ambiguity}
\label{sec:AdlerAmbiguity}

The potential linear divergence in a Class-B sub-diagram can be eliminated by adopting a symmetric integration scheme in which the contributions from integration regions with large positive and large negative values of the loop momenta cancel.  However, this choice of a symmetric integration scheme may introduce ambiguity.  Thus, the Class-B sub-diagrams evaluated in a lattice calculation may contain an unphysical component that depends upon the structure of the lattice action at short distances and how the operators in question are discretized.

Since sub-diagrams of Class-B become well-defined when differentiated twice with respect to the components of the external momenta or the quark masses, or both, the unphysical component that will require special treatment must be a local function of the $s$ and $d$ quark fields and two E\&M vector potentials $A_\mu(x)$ and $A_\nu(x)$ and a first-order polynomial in the external momenta and quark masses.  There is only one such term which we will refer to as the ``Adler ambiguity''
\begin{equation}
 \zeta\, (\overline{s}\gamma^\sigma\gamma^5 d) \epsilon^{\mu\nu\sigma\rho}\left\{(\partial^\rho A^\mu) A^\nu - A^\mu(\partial^\rho A^\nu)\right\}
 \label{eq:adler}
\end{equation}
since this term was explicitly identified in Eq. (A1) of Adler's remarkably complete discussion of the properties of the triangle graph and its axial anomaly~\cite{Adler:1969gk}.  The requirement that this ambiguous amplitude must vanish if differentiated twice with respect to the external momenta or quark masses implies that the coefficient $\zeta$ must be a constant.

The operator shown in Eq.~\eqref{eq:adler} is the unique dimension-6 operator which has the required Bose symmetry between the two emitted photons and a non-vanishing matrix element between a $K_L$ state and vacuum.  As can be seen from its structure, this operator would be forbidden in a lattice formulation with a conserved E\&M current.  However, it will appear in our $K_L\to\mu^+\mu^-$ calculation which uses the local, non-conserved current and it will generate an unphysical contribution to the resulting $K_L\to\mu^+\mu^-$ amplitude unless the operator in Eq.~\eqref{eq:adler} with a properly chosen coefficient $\zeta$ is added as a correction term to remove it.

The discussion above has focused on the simplest, one-loop Class-B sub-diagram.  If we work to order $\alpha_s(m_c)^0$, this is the only diagram that would require such a correction term.  However, as we add more gluon propagators the linear divergence and its associated ambiguity remain.  Were we explicitly discussing the axial anomaly, we could invoke a variety of non-renormalization theorems to argue that it is only the simplest, one-loop diagram that actually contributes, such as for the case of QED~\cite{Adler:1969er} or QCD~\cite{Costa:1977pd}.  However, we are not discussing the axial anomaly in the case of a conserved E\&M current with Feynman or dimensional regularization but a potential ambiguity that may appear in a lattice regularization which violates current conservation. This failure of conservation is visible only at the cut-off scale, a scale at which the effects of the regularization are substantial.  We are not discussing the topological charge which is invariant under the small fluctuations that underlie a perturbative expansion.  Thus, in the absence of a deeper understanding of the Adler ambiguity, if we choose to work beyond O$(\alpha_s(m_c)^0)$ we will not assume that it is only the simple triangle graph without gluon corrections that contributes to this ambiguity and instead treat all Class-B sub-diagrams as requiring a correction term of the form given in Eq.~\eqref{eq:adler}.  Fortunately, this correction term is not difficult to determine non-perturbatively.

\subsection{Resolving the three-flavor Adler ambiguity}
\label{sec:ambiguity-fix}

The correction term given in Eq.~\eqref{eq:adler} is present because at short distances the lattice E\&M current we are using is not conserved.  The correction term given in that equation also introduces a contribution to the E\&M current which is not conserved.  Thus, the natural criterion to apply to determine the unknown constant $\zeta$ is that each of the two E\&M currents, after the correction term has been added, is conserved. 

We define the corrected hadronic matrix element as
\begin{eqnarray}
R^{\mu\nu}(k_1,k_2,P_K,x) &=& \int d^4 u\; d^4v\; e^{-i(k_1u+k_2v)} \langle 0|T\left\{ J_\mu(u) J_\nu(v) \mathcal{H}^{\Delta S=1}(x)\right\}|K_L(\vec P_K)\rangle 
    \label{eq:adler-corrected}    \\
  && \hskip 0.8 in + \zeta\, e^{-i(k_1+k_2)x}\epsilon^{\mu\nu\sigma\rho}(k_1-k_2)^\rho \langle 0|(\overline{s}\gamma^\sigma\gamma^5 d)(x) |K_L(\vec P_K)\rangle, \nonumber
\end{eqnarray}
where for clarity we have evaluated both the weak Hamiltonian and the correction term at the non-zero position $x$.  The term on the right-hand side of the first line of Eq.~\eqref{eq:adler-corrected} is the usual hadronic matrix element that we must evaluate when computing $K_L\to\mu^+\mu^-$ decay where our usual non-leptonic kernel $K^{\mu\nu}(u,v)$ has been replaced by two plane waves.  The amplitude in the second line is the hadronic matrix element of the correction term which is proportional to the axial current matrix element determined by $f_K$.  We can then obtain a value for $\zeta$ by solving the linear equation:
\begin{equation}
    (k_1)_\mu R_{\mu\nu}(k_1,k_2,P_K,x) = 0 \label{eq:FindZeta}
\end{equation}
for any choice of the open index $\nu$.

Since the difficulty that the correction term is introduced to resolve is a short-distance ambiguity, a single choice of $\zeta$ will result in Eq.~\eqref{eq:FindZeta} being obeyed for all long-distance kinematics.  Thus, if a choice of $\zeta$ implies that Eq.~\eqref{eq:FindZeta} is true for one value of the free index $\nu$ it must be true for all values.  Similarly with this choice of $\zeta$ Eq.~\eqref{eq:FindZeta} should be true for all choices of the four-momenta $k_1$ and $k_2$ provided their components are small compared to the inverse lattice spacing.  

As both of the operators appearing in Eq.~\eqref{eq:adler-corrected} are exponentially localized around the position $x$, in the spirit of our QED$_\infty$ formalism we can treat the components of the two four-momenta $k_1$ and $k_2$ as continuous variables that need not obey a finite-volume quantization condition provided the position $x$ is chosen near the center of the periodic volume in which the lattice QCD calculation is performed.  All of these statements can be verified numerically and their lack of accuracy will provide an estimate of systematic error.

An important kinematic property of the condition given in Eq.~\eqref{eq:FindZeta} is the requirement that a non-zero momentum $q = k_1+k_2-P_K$ must flow into the operator densities $\mathcal{H}^{\Delta S=1}(x)$ and $\overline{s}\gamma^\sigma\gamma^5 d(x)$.  Since the matrix element of the axial current between the $K_L$ and vacuum states is proportional to the kaon momentum $P_K$, the term proportional to $\zeta$ in Eq.~\eqref{eq:adler-corrected} contains the factor:
\begin{equation}
\epsilon^{\mu\nu\rho\sigma}(k_1-k_2)^\rho P_K^\sigma.  \label{eq:adler_kinematics1}
\end{equation}
In the case that $P_K=k_1+k_2$ the quantity in Eq.~\eqref{eq:adler_kinematics1} becomes 
\begin{equation}
2\epsilon^{\mu\nu\rho\sigma}k_1^\rho\, k_2^\sigma.  \label{eq:adler_kinematics2}
\end{equation}
which vanishes when contracted with $k_1^\mu$ or $k_2^\mu$ implying that for these kinematics the condition Eq.~\eqref{eq:FindZeta} will be automatically obeyed and cannot be used to determine $\zeta$.

Injecting non-zero momentum into the four-quark operators appearing in $\mathcal{H}^{\Delta S=1}$ represents a well-defined extension of the effective three-flavor theory with one exception.  When the $u$ and $\overline{u}$ fields appearing in $Q_1$ and $Q_2$ are contracted a quadratic divergence is introduced whose renormalization required the introduction of lower-dimensional counter terms of the form $\overline{s}d$ and $\overline{s}\gamma^5d$.  
These counter terms are conventionally ignored because they are total divergences of the $\Delta S=1$ vector or axial vector currents and in lattice QCD calculations of decay rates the initial and final four momenta are the same so such total divergences will vanish.  This is no longer the case for the amplitudes we are introducing to determine $\zeta$.  We observe that this is not a source of difficulty here because, were they introduced in place of $\mathcal{H}^{\Delta S=1}$, such dimension-four counter terms would not contribute to Class-B sub-diagrams with degree of divergence +1 and would not enter a determination of $\zeta$.

We point out that Eq.~\eqref{eq:FindZeta} can be used consistently in an expansion in $\alpha_s(m_c)$.   If we are working to order $\alpha_s(m_c)^0$ then only the lattice QCD amplitude from the quark-line topology of Type-3 needs to be evaluated when calculating the two-current matrix element appearing in $R^{\mu\nu}$ and $\zeta$ will correctly include the $\alpha_s(m_c)^0$ term required for a physical result.   Since the calculation is non-perturbative, $\zeta$ will also include higher powers in $\alpha_s(m_c)$ which can be neglected at order $\alpha_s(m_c)^0$.  If we wish to work to order $\alpha_s(m_c)^1$ then, referring to Table~\ref{tab:sub-diag-contract}, we recognize that the quark line topologies of Type 1 and Type 4 must also be included when evaluating $R^{\mu\nu}$ and determining $\zeta$.  A calculation accurate to all orders in $\alpha_s(m_c)$, which should be possible, requires including all five quark line topologies when evaluating $R^{\mu\nu}$.

We conclude that the ambiguity introduced by the Class-B sub-diagrams can be resolved while working within the three-flavor theory by simply imposing current conservation to determine a single correction term, proportional to $f_K$ that must be added to the $K_L\to\mu^+\mu^-$ calculation.

\section{Class-C sub-diagrams}
\label{sec:Class-C}

The third class of sub-diagram that requires renormalization in our three-flavor calculation is the improper sub-diagram which corresponds to the entire graph shown in the second row of Figure~\ref{fig:subdiagrams}.  In a four-flavor theory with the approximation of Cabibbo unitarity the entire graph would be convergent and no counter term would be required.  However, in our three-flavor lattice calculation, the entire graph is divergent and we need to determine a counter term of the form
\begin{equation}
C_1^C (\overline{s}\gamma^\mu\gamma^5 d)(\overline{\mu}\gamma^\mu\gamma^5 \mu) \label{eq:Class-C-CT}
\end{equation}
with the coefficient $C_1^C$ chosen to remove the lattice-spacing dependent result from the three-flavor calculation and to replace it by the physical dependence on the charm quark mass. 

The coefficient $C_1^C$ can be determined non-perturbatively by comparing the evaluation of the $K_L\to\mu^+\mu^-$ decay amplitude for the quark line topologies of interest between three- and four-flavors, including the counter term in Eq.~\eqref{eq:Class-C-CT} in the three flavor calculation and adjusting $C_1^C$ to make the two results agree.

As in the case of the Class-A renormalization we can make the needed three- and four-flavor comparison for the contribution of specific quark line topologies to the $K_L\to\mu^+\mu^-$ decay amplitude using unphysical kinematics.  By choosing to use heavier-than-physical masses for the light quarks, we can suppress finite volume corrections on the relatively small volume of the 32IF ensemble, $\mathcal{E}_{N_f=4}$, that would be used for the four-flavor calculation.  The difficulties caused by intermediate states that are lighter than the kaon can also be eliminated if the $u$ or $d$ is given a mass heavier than the strange quark.  (These problems could also be avoided by working with a fixed maximum separation between the weak operator and the E\&M current closest to it.)  Except for these unphysical quark masses and the introduction of a charm quark in the four-flavor calculation, both the three- and four-flavor calculations can mirror the original $K_L\to\mu^+\mu^-$ calculation with an initial kaon created by a wall source and our standard non-hadronic kernel $K_{\mu\nu}$.  However, with heavier than physical light quark masses an accurate result would require substantially less calculation.  The final step would be a comparison of the unphysical results between the three- and four-flavor calculations allowing a value of $C_1^C$ to be determined that makes the two results agree.

As in the case of the Class-A and Class-B sub-diagrams, we can organize the calculation of the counter terms required by these Class-C diagrams as a power series in $\alpha_s(m_c)$.  As indicated in Table~\ref{tab:sub-diag-contract}, if we choose to work only to O$(\alpha_s(m_c)^0)$ then we would need to follow the steps above only for diagrams with Type-1 and Type-3 quark line topologies.  Working to O$(\alpha_s(m_c)^1)$ would require that Type-2 and Type-4 topologies are included while Type-5 topologies would enter at O$(\alpha_s(m_c)^2)$.

\section{Conclusion}\label{sec:conclu}

A lattice QCD calculation of the two-photon exchange contribution to $\kl\to\mu^+\mu^-$ decay is most accessible in a three-flavor theory since the inclusion of a charm quark requires the use of a small lattice spacing while the need to accurately treat the pion requires a large physical volume.  In this paper we have shown how to resolve the additional complications that arise in such a three-flavor theory from the absence of GIM cancellation.  A series of counter terms depending on low-energy constants is required for the three-flavor theory to be an effective  theory that accurately represents the better-defined four-flavor theory at low energies below the charm quark mass.  

The three-flavor lattice calculation is conventionally organized according to the various topologies of quark contractions that correspond to the products of quark propagators that must be evaluated by Monte Carlo averages.  In contrast, the identification of the necessary counter terms is based on the degree of divergence of sub-diagrams with specific numbers of quark and gluon external lines.  Here we have organized the relation between these two descriptions using a power series expansion in the strong coupling $\alpha_s$ evaluated at the charm quark mass $m_c$ since in the three-flavor calculation undertaken in Ref.~\cite{Boyle:2025fug} the lattice scale is close to $m_c$.  

In contrast with the typical weak interaction effective theory where the low energy constants must be determined from physics at the scale of the $W$ and $Z$ bosons and rely on QCD perturbation theory, the low energy constants appearing in the three-flavor theory can be determined non-perturbatively from the four-flavor effective theory.  Because these three-flavor counter terms correct the short distance properties of the three-flavor theory they are independent of the masses of the $u$, $d$ and $s$ quarks.  This allows the three- to four-flavor comparisons which determine those constants to carried out economically, even if done non-perturbatively, by using larger than physical quark masses and consequently smaller that usual physical volumes.  Thus, if desired, one can work to all orders in $\alpha_s(m_c)$.

The next step in determining the two-photon exchange contribution to $\kl\to\mu^+\mu^-$ decay is to use the methods developed in this paper to renormalize the three-flavor calculation reported in Ref.~\cite{Boyle:2025fug}, obtaining a complete result for a single lattice spacing.

\section*{Acknowledgments}
We thank our RBC and UKQCD Collaboration colleagues for discussion and ideas.
E.-H.C. would like to thank Marc Knecht for illuminating discussion.
E.-H.C. was supported by the U.S. Department of Energy, Office of Science, Office of Nuclear Physics under grant Contract Number DE-SC0011090.
N.H.C. was supported in part by the U.S. Department of Energy (DOE) grant No.~DE-SC0011941. 

\newpage

\appendix

\section{Analytic properties of the leptonic kernel}\label{sec:sd-kernel}

In this Appendix, we provide an alternative derivation of the coordinate-space leptonic kernel that enters the calculation of the dispersive LD2$\gamma$ amplitude.
At the end, a more compact expression than that obtained in Ref.~\cite{Chao:2024vvl} is given, which makes it easier to study the analytic properties of the kernel.
Unless otherwise stated, we work in Minkowski space with mostly-negative metric convention and with space-like separations.
An analytic continuation $(r^0,\vec{r})\to(-ir^0,\vec{r})$ is eventually needed for lattice applications.

The central quantity that we need for the dispersive LD$2\gamma$ leptonic kernel is the tensor
\begin{equation}\label{eq:genker}
\mathcal{K}_{\mu\nu;\alpha\beta}(r;k^+) \equiv\left[ \gamma_\nu
\left(\frac{1}{2}\slashed{P}-i\slashed{\partial}+m_\mu\right)\gamma_\mu\right]_{\alpha\beta}\textrm{Disp}\;\mathcal{L}(r;k^+,k^-)\,,
\end{equation}
with $\textrm{Disp}\;\mathcal{L}(r; k^+)$ being the dispersive (real) part of  
\begin{equation}\label{eq:ldef}
\mathcal{L}(r;k^+,k^-) = i\int\frac{d^4 q }{(2\pi)^4}\frac{e^{-iq\cdot r}}{\left(\frac{P}{2}-k^+ - q\right)^2 - m_\mu^2 + i\varepsilon}
\frac{1}{\left( \frac{P}{2} + q\right)^2 + i\varepsilon}
\frac{1}{\left(\frac{P}{2}-q\right)^2 + i\varepsilon}\,,
\end{equation}
where $\alpha$ and $\beta$ are the spinor indices, $m_\mu$ is the mass of the muon, $k^\pm$ are generic (ie. not required to be on-shell) four-momenta for $\mu^\pm$ and $P\equiv k^++k^-$.

Denote 
\begin{equation}
R \equiv \frac{P}{2} - k^+\,,\quad
Q \equiv \frac{P}{2}\,.
\end{equation}
We can rewrite Eq.~\eqref{eq:ldef} as
\begin{equation}\label{eq:lcalfeyn}
\begin{split}
- i \mathcal{L}(r;k^+) = & 2 \int\frac{d^4q}{(2\pi)^4}\int\mathcal{D}[a,b] \frac{e^{-i(q+\mu)r}}{[q^2-\mu^2+a (R^2-m_\mu^2)+(1-a)Q^2 + i\varepsilon]^3}
\\ = &
2 \int\frac{d^4q}{(2\pi)^4}\int\mathcal{D}[a,b] e^{-i\mu r}\left[\frac{e^{-iqr}}{(q^2-\sigma + i\varepsilon)^3}\right]
\,,
\end{split}
\end{equation}
where
\begin{equation}
\mu \equiv a R - (1-a-2b)Q\,,\quad
\int\mathcal{D}[a,b]\equiv \int^1_0d a\int^1_0d b\;\Theta(1-a-b)\,,
\end{equation}
\begin{equation}
\sigma \equiv - a(1-a)R^2 + [(1-a-2b)^2-(1-a)]Q^2 
-2a(1-a-2b) R\cdot Q
+ a m_\mu^2\,.
\end{equation}
As the dispersive part comes from the contribution of the Hadamard finite part of the quantity in the square-bracket, the quantity is well-regulated and we can swap the order between the integrals over $a$, $b$ and $q$ while computing the dispersive part kernel. 
Using the expression for the Feynman propagator with a spacelike separation $r$ 
\begin{equation}\label{eq:gfpos}
G_{\rm F}(\tau;m) = i\int\frac{d^4p}{(2\pi)^4}\frac{e^{-ip\cdot r}}{p^2-m^2+i\varepsilon} = 
\frac{m}{4\pi^2\tau}K_1\left(m\tau\right)\,,
\quad \tau \equiv \sqrt{-r^2-i\varepsilon}\,,
\end{equation}
we have
\begin{equation}\label{eq:lcal-master}
\textrm{Disp}\;\mathcal{L}(r;k^+,k^-) = \int\mathcal{D}[a,b] e^{-i\mu r} \textrm{Re}\left[\frac{\partial^2}{\partial \sigma^2}G_{\rm F}(\tau;\sqrt{\sigma-i\varepsilon})\right]\,.
\end{equation}
with
\begin{equation}
\begin{split}
& \textrm{Re} \frac{\partial^2}{\partial \sigma^2} G_{\rm F}(r;\sqrt{\sigma-i\varepsilon})=
\\&
\begin{cases}
\begin{split}
\frac{1}{64\pi^2\sigma^{3/2}r}
\Big[
& -2\sqrt{\sigma} r K_0(\sqrt{\sigma} r) +(-4+3\sigma r^2)K_1(\sqrt{\sigma}r)
\\& - 2\sqrt{\sigma} rK_2(\sqrt{\sigma} r) + \sigma r^2K_3(\sqrt{\sigma}r)
\Big]\,,
\quad\textrm{for }\sigma > 0 \,,
\end{split}
\\
\begin{split}
-\frac{1}{128\pi(-\sigma)^{3/2}r}
\Big[
&2\sqrt{-\sigma}r Y_0(\sqrt{-\sigma}r) + (-4+3\sigma r^2)Y_1(\sqrt{-\sigma}r) 
\\&- 2\sqrt{-\sigma} rY_2(\sqrt{-\sigma}r)- \sigma r^2 Y_3(\sqrt{-\sigma} r)
\Big]\,,
\quad\textrm{for }\sigma \leq 0
\,.
\end{split}
\end{cases}
\end{split}
\end{equation}
Here $Y_i$ and $K_i$ are Bessel functions and modified Bessel functions of the second kind respectively.

\paragraph*{The small-$r$ behavior of the dispersive LD2$\gamma$ kernel}

We can study the $r\to 0$ behavior of the dispersive LD2$\gamma$ leptonic kernel using the above expressions by projecting to the on-shell muon-antimuon state.
Placing the system in the lab frame, we arrive at a more compact expression for the real part of the leptonic kernel given in Eq.~(20) of Ref.~\cite{Chao:2024vvl}, up to the sign convention that is not kept track of,
\begin{equation}
\textrm{Re}L_{\mu\nu}(r) = 2m_\mu e^4 \varepsilon_{0\mu\nu\kappa} \frac{r^\kappa}{|\vec{r}|}\frac{\partial}{\partial |\vec{r}|}\bar{\mathcal{L}}(r;k^+,k^-)\,,
\end{equation}
where
\begin{equation}
\bar{\mathcal{L}}(r)\equiv\int d\hat{\Omega}_{\vec{k}^+}\textrm{Disp} \mathcal{L}(r;k^+,k^-)\,,
\end{equation}
is averaged over the solid angle of $\vec{k}^+$ to project to the ${}^1S_0$ $\mu^+\mu^-$-state.

We remark that the equal-time limit $r^0\to 0$ can be taken safely and we get for $|\vec{r}|\to 0$,
\begin{equation}
\textrm{Re}L_{\mu\nu}(r) = \frac{m_\ell e^4}{16\pi^2} \epsilon_{0\mu\nu\kappa} r^\kappa \ln|\vec{r}| + \mathrm{O}(|\vec{r}|) \,. 
\end{equation}
Thus, the dispersive kernel is smooth as $r^2\to 0$ but its first derivative diverges logarithmically.

\section{Reconstruction of a five-dimensional surface-to-bulk propagator from a four-dimensional physical propagator in the domain-wall fermion formalism}\label{sec:4Dto5D}

The five-dimensional surface-to-bulk M\"obius Domain-Wall Fermion (MDWF) propagator can be reconstructed from its four-dimension-projected physical propagator~\cite{Brower:2012vk}.
In this Appendix, we describe how to build the correlation functions with conserved currents given in Section~\ref{sec:Class-A} from a four-dimensional point-source DWF propagator. 

We begin by reviewing the MDWF action and fermion field with the notations of Ref.\cite{RBC:2014ntl}
\begin{eqnarray}
    S^5 &=& \bar{\psi}(x)[D^5_{GDW}(m)](x;y)\psi(y),\\
    Q_s(x)&=& \left[ \mathscr{P}^{-1}\psi\right]_s(x),\\
    q(x) &\equiv& Q_1(x),
\end{eqnarray}
where $[D^5_{GDW}(m)](x;y)$ is the five-dimensional MDWF Dirac operator with quark mass $m$, $\mathscr{P}$ is a five-dimensional matrix projecting onto left/right-handed component of fermion field on neighboring fifth dimension slices, $Q_s(x)$ denotes the fermion field at fifth dimension slice $s$ and four-dimensional coordinate $x$, and its first component corresponds to the four-dimensional physical fermion field $q(x)$.  

With this notation, the surface-to-surface propagator reads
\begin{eqnarray}
    \left<q(x)\bar{q}(y)\right> &=& \Tilde{D}^{-1}_{ov}(x;y)\nonumber\\
                                &=&\frac{1}{1-m}\left[D^{-1}_{ov}-1\right](x;y)\nonumber\\
                                &=&\frac{1}{1-m} \left[\mathscr{P}^{-1}D^5_{GDW}(m)^{-1}D^5_{GDW}(1)\mathscr{P}-1\right]_{11}(x;y)\label{eq:4d},
\end{eqnarray}
where $\Tilde{D}_{ov}(x;y)$ and $D_{ov}(x;y)$ represent the four-dimensional renormalized and bare overlap operators, related by
\begin{eqnarray}
    D^{-1}_{ov} = (1-m) \Tilde{D}^{-1}_{ov}  + 1\label{eq:re_to_ba}.
\end{eqnarray}
The subscript denotes the corresponding component of the five-dimensional matrix. Henceforth, the four-dimensional coordinates will be suppressed for simplicity. 

On the other hand, the five-dimensional surface-to-bulk propagator is given by~\cite{Brower:2012vk}
\begin{eqnarray}
    \left<Q_s \bar{q}\right> &=& \frac{1}{1-m} \left[\mathscr{P}^{-1}D^5_{GDW}(m)^{-1}D^5_{GDW}(1)\mathscr{P}-1\right]_{s1}\label{eq:stb}\\
    &=&
    \begin{pmatrix}
    \Tilde{D}^{-1}_{ov}\, \\
    \Delta^{R}_2\,D^{-1}_{ov}\,-\frac{1}{1-m} \\
    \vdots \\
    \Delta^{R}_{L_s}\,D^{-1}_{ov}\,-\frac{1}{1-m}
    \end{pmatrix},\\
    \Delta^R_{s+1} &=& \frac{T^{-(L_s-s)}}{1+T^{-L_s}}.
\end{eqnarray}
where $L_s$ is the extent of fifth dimension and $T^{-1}$ is the transfer matrix along the fifth dimension.  

Let $\eta$ be a point-source vector living on the four-dimensional surface and $\tilde{D}_{ov}^{-1}\eta$ be the corresponding renormalized physically-projected point-source propagator.
The surface-to-bulk propagator associated with the same source vector $\langle Q_s \bar{q}\rangle\eta$ can be reconstructed from the available $\tilde{D}_{ov}^{-1}\eta$ data with the mere cost of one inversion of the Pauli-Villars term, $D_{GDW}^5(m=1)$, thanks to Eq.~\eqref{eq:re_to_ba} and the relationships
\begin{eqnarray}
\phi \equiv \langle Q_s\bar{q}\rangle\eta
=\frac{1}{1-m}
\begin{pmatrix}
(1-m)\Tilde{D}^{-1}_{ov}\,\eta \\
\psi_2-\eta\\
\vdots \\
\psi_{L_s}-\eta
\end{pmatrix},\label{eq:final}
\end{eqnarray}
where $\psi_i$ denotes the $i$-th component of the five-dimensional vector $\psi$ defined via
\begin{eqnarray}
 \psi 
 &=&  \mathscr{P}^{-1}D^{5}_{GDW}(1)^{-1}D^{5}_{GDW}(m)\, \mathscr{P}\, \xi ,\label{eq:PV}\\
 \xi &=& 
 \begin{pmatrix}
-D_{ov}^{-1}\eta \\
0\\
\vdots \\
0
\end{pmatrix}\,.
\label{eq:sol}
\end{eqnarray}

The same technique is exploited by the M\"obius Accelerated Domain Wall Fermion (MADWF) algorithm, which obtains the five-dimensional surface-to-bulk DWF propagator by solving an approximate bare overlap propagator constructed from MDWF with smaller $L_s$ \cite{Yin:2011np}.

\bibliography{references}

\end{document}